\documentclass[12pt,preprint]{aastex} 

\newcommand{\Msun}{${M_{\odot}}$}

\newcommand{\am}{\arcmin}

\slugcomment{\today}

\shorttitle{Young Stars in the Carina Nebula}
\shortauthors{Sanchawala et al.}

\begin{document}

\title{NEAR-INFRARED STUDY OF THE CARINA NEBULA}


\author{
        Kaushar Sanchawala\altaffilmark{},
        Wen-Ping Chen\altaffilmark{}}
        
\affil{National Central University, Taiwan}

\author{
        Devendra Ojha\altaffilmark{},
        Swarna Kanti Ghosh\ \altaffilmark{}}
\affil{Tata Institute of Fundamental Research, India}

\author{
        Yasushi Nakajima\altaffilmark{}, 
        Motohide Tamura\altaffilmark{}}
\affil{National Astronomical Observatory of Japan, Japan}

\author{
        Daisuke Baba\altaffilmark{},
	Shuji Sato\altaffilmark{}}
\affil{Department of Astrophysics, Nagoya University, Japan}

\and

\author{
	Masahiro Tsujimoto\altaffilmark{}}
\affil{Pennsylvania State University, USA}



\begin{abstract}

We have carried out near-infrared (NIR) imaging observations of the Carina Nebula 
for an area of $\sim$400 arcmin$^{2}$ including the star clusters Trumpler\,14 (Tr\,14) and Trumpler\,16 (Tr\,16). With 10 $\sigma$ limiting magnitudes of $J \sim$18.5, $H \sim$17.5 and $K_s \sim$16.5, we identified 544 Class II and 11 Class I young star candidates. We find some 40 previously unknown very red sources with $H-K_s>2$, most of which remain undetected at the $J$ band. These young star candidates provide a comprehensive sample to diagnose the star-formation history of this massive star-forming region. The red NIR sources are found to be concentrated to the south-east of Tr\,16, along the `V' shaped dust lane, where the next generation of stars seems to be forming. 
In addition, we find indications of ongoing star formation near the three MSX point sources, G287.51-0.49, G287.47-0.54, and G287.63-0.72. A handful of red NIR sources are seen to populate around each of these MSX sources. Apart from this, we identified two hard $Chandra$ X-ray sources near G287.47-0.54, one of which does not have an NIR counterpart and may be associated with a Class I/Class 0 object. The majority of 
the Class II candidates, on the other hand, are seen to be distributed in the directions of the clusters, demarcating different evolutionary stages in this massive star-forming region. A comparison of the color-magnitude diagrams of the clusters with pre--main-sequence (PMS)
model tracks shows that the stellar population of these clusters is very young ($\lesssim$ 3 Myr). The $K_s$-band luminosity function (KLF) of Tr\,14 shows structure at the faint end, including a sharp peak due to the onset of deuterium burning, implying an age of 1--2 Myr for the cluster. The KLF of Tr\,16, in contrast, is found to rise smoothly until it turns over. The slopes of the mass functions derived for the clusters are found to be in agreement with the canonical value of the field star initial mass function (IMF) derived by \citet{salpeter55}.

\end{abstract}





\section{INTRODUCTION}

The Carina Nebula (NGC\,3372) is one of the most active 
star-forming regions of the Milky Way Galaxy harboring more than a dozen 
star clusters \citep{feinstein95}. Trumpler\,14 and Trumpler\,16 (Tr\,14 and Tr\,16 hereafter), located in the central part of the 
Nebula, are the youngest and the most populous clusters of this region. There are at least 31 O type stars
in these two clusters \citep{feinstein95}, with six exceedingly rare O3 type main-sequence stars.
Tr\,16 hosts $\eta$ Carinae, which is a typical luminous blue variable and arguably the most massive star 
of the Galaxy \citep{mj93}.
There are also three Wolf-Rayet (WR) stars in this region, HD\,93162, HD\,92740 and HD\,93131, of which, HD\,93162 is 
most certainly a member of Tr\,16, whereas the other two are likely members of Tr\,16 but are located about 25$\arcmin$ to the
west and south of $\eta$ Carinae respectively \citep{walborn95}. 

These clusters, owing to their rich stellar contents, are extensively studied in different wavelengths. For instance, \citet{mj93} presented optical photometry
and spectroscopy of the brightest and bluest stars of the clusters. \citet{cudworth93} carried out a proper-motion study of
nearly 600 stars of the clusters with a limiting V magnitude of 15.5.  Recently, a multiwavelength imaging study 
of the clusters was carried out by \citet{tapia03}, in which they studied the cluster properties and estimated the ages of these clusters to be between 1--6 million years. The molecular cloud observations of the Nebula
have been made by several groups. Two large CO emission regions are found, known as the northern cloud and the southern cloud
 \citep{degraauw,whiteoak},
both of which are believed to be a part of a much larger Carina molecular cloud complex with a mass in excess of 5$\times$10$^5$ \Msun~ and
a projected length of 130 pc \citep{grabelsky}. The molecular gas
in the vicinity of Tr\,16 has been photoevaporated and a large cavity can be seen in the northern molecular 
cloud \citep{cox95,brooks98}, whereas a strong interaction seems to be occurring between Tr\,14 and the northern molecular cloud. 
As for the distribution of the ionized gas, a large ionized region is found with two concentrations, known as Car I and Car II, located to the west of Tr\,14, and to the north of $\eta$ Carinae respectively \citep{gardner,brooks01}. 

The reported distance modulus for Tr\,16 ranges from 11.8 \citep{levato81} to
12.55 \citep{mj93}, and for Tr\,14, from 12.20 \citep{feinstein83} to 12.99 \citep{morrell88}. It has been shown by \citet{turner80} and \citet{mj93} that the two clusters are located at the same distance from the sun, whereas some groups have found that they are located at slightly different distances \citep{walborn73,morrell88}. For our study, we have assumed a common distance of 2.5 kpc for both the clusters.
  
The picture emerging of this massive star-forming region in the last decade suggests that the star formation
activity is still ongoing. The first evidence of star formation activity came out in the study by \citet{megeath}, in which they found an object,
IRAS\,10430-5931, to be associated with a bright rimmed globule to the south-east of Tr\,16, along the thick `V' shaped dust lane that bisects the
Carina Nebula. Using the mid-infrared observations from MSX (Midcourse Space Experiment), and optical emission line observations, \citet{smith00}
found several clumps along the edge of the dark cloud and to the east of $\eta$ Carinae, including the clump associated with
IRAS\,10430-5931. They remarked that these IR globules which lie behind the ionization fronts, in the periphery of the Nebula, 
are candidate sites of triggered star formation. In another case, \citet{smith03} identified numerous protoplanetary disk (proplyd) candidates in this region. Moreover, widespread photodissociation regions (PDRs) are also found throughout the Nebula on the surfaces of molecular clouds \citep{rathborne02}. 
In our earlier work on this region based on the $Chandra$ data \citep[Paper I hereafter]{sanchawala}, we found a compact ($\sim$2 pc across, assuming a distance of 2.5~kpc)
group of ten X-ray sources, located close ($\sim$4\arcmin) to IRAS 10430-5931 and along the `V' shaped dust lane. This compact X-ray group is referred to as the Tr\,16-SE group hereafter. More than half of the X-ray sources of the Tr\,16-SE group 
are massive star candidates as judged by their X-ray and NIR properties (see Paper I). All the above examples seem to trace the current generation of star formation which is possibly triggered by the earlier generation of cluster stars [see, e.g. \citet{chen07} for examples of triggered star formation in OB associations].

In this paper, we present deep NIR imaging observations ($JHK_s$) covering $\sim$400 arcmin$^{2}$ area, including the clusters Tr\,14 and Tr\,16.
To aid and complement the identification of young stellar populations of the region, we made use of the X-ray sample of the region derived from the $Chandra$ archival datasets, ObsIDs=1249, and 50 (Paper I), and ObsID=4495 (unpublished). For the $JHK_s$ color-composite image, readers are referred to Paper I. Though NIR imaging observations of Tr\,14 and Tr\,16 have been reported earlier, for example the recent work by \citet{tapia03}, our observations are deeper by 2--3 magnitudes than any existing NIR observations.
Moreover, we produce the most comprehensive sample of low-mass PMS candidates including 
Class I and Class II candidates identified based on the NIR colors, as well as Class II and Class III candidates identified based on the NIR and X-ray properties, as the first step to study how star formation proceeds in violent environments and the role massive stars play in the ongoing star-formation activity. The spatial distribution of sources suggests that star formation is in different evolutionary phases in this region with Tr\,16 being the most evolved population. Tr\,14 seems to be younger than Tr\,16 based on the relative fraction of sources with NIR excesses. We construct the $K_s$-band luminosity functions (KLFs) of the clusters to constrain the ages and initial mass functions (IMFs) of the clusters. The Tr\,16-SE group shows a clear excess of sources relative to the field and thus should be an embedded stellar group or cluster. We also find strong indications of ongoing star formation near the three MSX sources, G287.51-0.49, G287.47-0.54, and G287.63-0.72. An approximate estimate of the initial mass functions (IMFs) of the clusters is also presented.

The paper is organized as follows. In \S{2} we present the NIR observations carried out at the InfraRed Survey Facility (IRSF), the data reduction procedures for the IRSF observations and the description of other datasets used in this work, i.e., 2MASS and $Chandra$. In \S{3} we present the results and discussions of our work and \S{4} summarizes the paper.

\section{OBSERVATIONS AND DATA REDUCTION} 

\subsection{InfraRed Survey Facility (IRSF)}

\subsubsection{Observations}

The NIR imaging observations toward the Carina Nebula were carried out on 
2003 April 1, 2, 9, and 10 using the SIRIUS (Simultaneous InfraRed Imager for Unbiased Survey) camera mounted on the IRSF 1.4 m telescope in
Sutherland, South Africa. The camera is equipped
with three HAWAII arrays, each of $1024 \times 1024$ pixels and dichroic mirrors enabling simultaneous observations in the  $J$ (1.25 $\mathrm{\mu m}$), $H$ (1.63 $\mathrm{\mu m}$) and $K_s$ (2.14 $\mathrm{\mu m}$) wavebands.
It provides a plate scale of 0\farcs45 $\mathrm{pixel^{-1}}$
for a field of view of 7\farcm8 $\times$ 7\farcm8. More details about the camera are discussed in \citet{nagayama}. 
We observed nine fields 
toward the Nebula (see Table~1 for the observation log) in 3$\times$3 mosaics, with an overlap of
$\sim$1$\arcmin$ between the frames in both directions. The composite observed field is centered on
the coordinates of $\eta$ Carinae, i.e., R.\,A. = $10^h45^m05^s$ and Decl. = $-59\arcdeg 38\arcmin 52\arcsec$ (J2000.0), 
covering a total area of $\sim$$20\arcmin \times 20\arcmin$. 
For each pointing, 30 dithered frames were observed, each with an integration time of 30~s, giving a total
integration time of 900~s.  Two pointings (\#5 and \#6; see Table~1) observed on April 9, 2003
were affected by bad weather, so were re-observed on 2005 January 14.
For these later observations, 45 dithered frames were observed, each with an 
integration time of 20~s, yielding again a total integration time of 900~s for 
each pointing. A reference field centered on R.\,A.=$10^{h}47^{m}17.4^{s}$ and Decl.=$-59^{\arcdeg}39^{\arcmin}26^{\arcsec}$ (J2000) was observed with the same integration time as for the target fields. The typical seeing (FWHM) during our observations ranged from 
1\farcs0 to 1\farcs4 and the airmass from 1.2 to 1.5.  For photometric calibration, the standard stars No.~9144 and 9146 from \citet{persson} were observed on each night.

\subsubsection{Data Reduction}

We used the NOAO's IRAF\footnote{IRAF 
(the Image Reduction and Analysis Facility) is distributed by the National Optical
Astronomy Observatory, which is operated by the Association of Universities for 
Research in Astronomy, Inc., under cooperative agreement with the National Science Foundation} package to
reduce the IRSF data. The standard procedures for image reduction were 
applied, including dark subtraction, sky subtraction and flat field
correction.  The dithered images in each band were then combined 
for each pointing to achieve a higher signal-to-noise ratio. We performed
photometry on the reduced images using the IRAF's DAOPHOT package
\citep{stetson}. Since Carina Nebula is a very crowded region and with a large
amount of diffuse emission in the NIR, the point spread function (PSF) photometry was performed. To
construct the PSF for a given image, we chose about 10 bright stars well
isolated from neighboring stars and located away from nebulosities or 
edges of an image. The task ALLSTAR of DAOPHOT was then used to apply the average PSF to all the stars in the image, 
from which the instrumental magnitude of each star was derived. The
instrumental magnitudes were then calibrated against the standard stars observed on each night.
The magnitudes we derived are in the SIRIUS color system. For the purpose of plotting this data in the
color-color and color-magnitude diagrams, we have converted our magnitudes to the CIT system using the color conversion equations between the SIRIUS and CIT systems derived by
\citet{nakajima}.

To remove spurious detections from our catalogues, the IRAF parameters {\it sharpness} and {\it chi} of the ALLSTAR task were used. 
These parameters measure the roundness of the object and the goodness of the PSF fit, respectively. Empirically, we found that with 
a cut at {\it sharpness}=1.5, and {\it chi}=4, we could eliminate most of the spurious detections and hence reduce much of the scatter in the 
magnitude error plot.
 
Since we performed photometry on individual pointings, the catalogues of all the pointings were merged in the end to produce the master catalogues of the composite field in the $J$, $H$, and $K_s$-bands. For sources
lying in the overlapping regions, the magnitudes varied by up to $\sim$0.1 mag. For these sources the average value of the magnitude was adopted.

We have obtained photometry for $\sim$41,000 sources in the $J$ band, $\sim$50,000 sources in the $H$ band and $\sim$36,000 sources in the $K_s$ band. Sources are saturated at $K_s<$ 11 mag.  The limiting magnitudes at 10$\sigma$ are
estimated to be $\sim$18.5, 17.5, and 16.5, in the $J$, $H$, and $K_s$ bands, respectively. 
We estimated the completeness limits of our data by adding artificial stars in our images using the ADDSTAR package in IRAF. These artificial stars were added at a 0.5 magnitude interval and at random places in the images. The detection rate was then derived
as a function of magnitude by calculating the fraction of the artificial stars that were recovered in the photometric analysis. This procedure was repeated five times.  
The 90\% completeness limits of our data for the $J$ band varied from 15.0 to 15.7 mag, for the $H$ band from 14.7 to 15.2 mag, and for the $K_s$ band from 14.7 to 15.3 mag, 
though we detect sources by up to 3 magnitudes fainter in all the bands.
For comparison, the 90\% completeness limits for the reference field are reached at 17.25, 16.25, and 16.75 mag in the $J$, $H$, and $K_s$ bands, respectively. The higher number density of sources in the clusters
as compared to the reference field, and the non-uniform diffuse NIR emission in the Nebula, are the reasons for shallower completeness limits for the clusters.

By using WCStools\footnote{http://tdc-www.harvard.edu/software/wcstools/}, we converted our pixel coordinates to R.\,A. and Decl. coordinates. The USNO-A2.0
catalogue was utilized to select some common sources in the frames to get the transformation relation. 
The astrometric accuracy is estimated to be $\sim$0.5$\arcsec$.

\subsection{2MASS}

Since sources brighter than $K_s < $ 11 mag are saturated in our IRSF data, we made use of the 
2MASS (Two Micron All Sky Survey)\footnote{This publication makes use of the data products from the Two Micron All Sky Survey, which is a joint project of the University of Massachusetts and the Infrared Processing and Analysis Center/California Institute of Technology, funded by the NASA and the NSF.} database to study the bright-end of the stellar population. We downloaded the
NIR sources from the 2MASS database within our observed IRSF field, with a certain constraints in order to select sources only with good photometric quality. 
For this, we used the 2MASS flags, namely, read flag, 'Rflg', blend flag, 'Bflg', and photometric quality flag, 'Qflg'. The 'Rflg' and 'Bflg' indicate the quality of photometry and the blending of a source, respectively, whereas the 'Qflg' indicates the signal-to-noise ratio of the detection. For our source selection, we considered values of 1--3 for 'Rflag', 1 for 'Bflag', and 'A' for 'Qflg'. The 2MASS magnitudes were also converted to the CIT system\footnote{http://www.astro.caltech.edu/~jmc/2mass/v3/transformations/} before combining with our IRSF sample.

\subsection{$Chandra$}

To complement our NIR data, we use the X-ray sample of the Carina Nebula derived from the multiple datasets of the archival \textit{Chandra X-ray Observatory} \citep{weisskopf02} data.  A sample of 450 X-ray sources derived from the merged datasets, ObsIDs=1249, and 50, observed by the Advanced CCD Imaging Spectrometer (ACIS--I) \citep{garmire03} in September 1999, was used (Paper I). The merged observation had a net exposure time of 18 ks, giving a limiting X-ray flux of $\sim$10$^{-14}$ ergs cm$^{-2}$ s$^{-1}$. Deeper observations of Tr\,14 region were carried out by the $Chandra$ ACIS--I in September 2004. We analyzed this recent dataset, ObsID 4495, that has an exposure time of 60 ks. An energy range of 0.5--8.0 keV was adopted for the detection of sources. The CIAO \citep{fruscione06} program WAVDETECT was used to detect the sources, with the wavelet scales ranging from 1 to 16 pixels in steps of $\sqrt{2}$, and a source significance threshold of 3 $ \times 10^{-6}$. The X-ray counts were estimated using the DMEXTRACT tool of CIAO. For more details of data analysis procedure, the reader is referred to Paper I. We extracted in total some 500 X-ray sources, 150 of which are not covered by the IRSF field. Together with the previous sample of 450 X-ray sources, our combined sample of the region has 800 sources.

\section{RESULTS AND DISCUSSIONS}

Figure 1 shows the mosaiced $K_s$ band image, centered on $\eta$ 
Carinae, with the north to the top and east to the left (see also Fig.~2 of Paper I for the $JHK_s$ color-composite image of the same field). Tr\,16 is seen near the center in the image, whereas Tr\,14 is seen in the north-west. The  WR star, HD 93162 (WR 25), is marked in the figure. Also marked are three MSX point sources, G287.51-0.49, G287.47-0.54, and G287.63-0.72, which are candidate embedded clusters \citep{rathborne}. The IRAS source, 
IRAS\,10430-5923, is known to have IRAS colors characteristic of an embedded star \citep{megeath}. In addition to this, we found three more IRAS sources in our field, IRAS 10424-5916, IRAS 10439-5927, and IRAS 10419-5925, which have IRAS colors characteristic of an embedded star.
All of these IRAS sources are marked in the figure. Finally, the Tr\,16-SE group is also marked in Fig. 1.

A large amount of diffuse NIR emission is seen in the field, extending northwest of $\eta$ Carinae, which 
probably arises due to both free-free and bound-free emission. The diffuse emission correlates very well with the 4.8-GHz continuum emission from Car II (see Fig. 3 in \citet{brooks01}). To the south-east of Tr\,16, where a thick `V' shaped dust lane bisects the Carina Nebula, some bright rimmed clouds or globules are seen, pointing toward $\eta$ Carinae or to Tr\,16. The most prominent of these is the bright rimmed globule seen nearly 7$\arcmin$ south of $\eta$ Carinae, which was found by \citet{megeath} to be associated with IRAS\,10430-5931 (see Fig.~1). CO emission has also been detected from this region \citep{brooks98}, which traces the northern edge of the southern molecular cloud.

We selected the sources detected in all the $J$, $H$, and $K_s$ bands using a search radius of 1$\arcsec$. There are 27,256 sources common in the $J$, $H$, and $K_s$ bands in the complete field. Of these, 17,495 sources have photometric errors
smaller than 0.1 mag in all the three bands. From the 2MASS data, we had an additional 226 bright sources with $K_s<11$ mag. To search for the X-ray counterparts of the NIR sources, our X-ray sample based on the multiple $Chandra$ datasets (see \S{2.3}) was utilized. With a progressively large search radius criteria of 1--3$\arcsec$ depending upon the off-axis angle of the X-ray source, we found counterparts of some 600 NIR sources.
To find the counterparts of known OB stars, the sample of known OB stars from \citet{mj93} was used. Within a search radius of 2$\arcsec$, our search resulted in 50 O to early-B stars.

\subsection{Color-Color Diagrams}

Figure 2a shows the color-color diagram of the 17,495 sources from the IRSF data (with photometric errors smaller than 0.1 mag) and of 226 bright NIR sources from 2MASS.
The color-color plots for the NIR sources with counterparts of known OB stars, and with X-ray counterparts, are shown separately in Fig.~2b and Fig.~2c. The 
intrinsic colors of dwarfs and giants are plotted as solid and dashed
line, respectively, which are taken from \citet{bb} after being converted to the CIT system. The two long parallel dashed lines adopted as $A_J/A_V = 0.265$, $A_H/A_V = 0.155$, and $A_K/A_V = 0.090$ \citep{cohen81}, drawn from the base and tip of the dwarf and giant loci, form the reddening band. The crosses marked on the reddening vectors
are separated by $A_V$ = 5 mag. The dotted line represents the locus of unreddened classical T Tauri stars \citep{meyer}. 
One more reddening vector is plotted starting at the tip of the CTTS locus.
The region bounded by the short dashed lines below the CTTS locus is where the PMS stars of intermediate mass, i.e., Herbig Ae/Be stars are 
usually found \citep{hernandez}. 

As shown in Fig.~2a, the NIR sources are classified in three regions
[see, e.g., \citet{tamura98} and \citet{ojha04}]. ``F'' sources are mostly reddened field stars although PMS stars with little NIR excesses, viz., weak-lined T Tauri sources or CTTSs may also be included. The sources located redward of region ``F'' have NIR excesses. Among these, ``T'' sources show small amounts of NIR excesses. These are T Tauri or Class II objects \citep{lada92}. It should be noted that this selection would only give a lower limit on the number of T Tauri objects in the sample, as T Tauri objects with small amounts of NIR excesses may well be located in the reddening band and those not highly reddened would be located near the CTTS locus. Sources which fall redward of the region ``T'' show large amounts of NIR excesses. These ``P'' sources are protostar-like (also known as Class I) objects. It should be noted that Herbig Ae/Be stars overlap partially with Class I objects in NIR colors. 
Based on these criteria, we found 544 Class II candidates and 11 Class I candidates.

For comparison, the color-color plot of the reference field is shown in Figure 3. 
The dwarf and giant loci, the CTTS locus, as well as the reddening bands are plotted in the same way as in Fig.~2. The area of the reference field is 19.36 arcmin$^{2}$. There are 402 sources detected in this field which are common to all the three bands and have photometric errors smaller than 0.1 mag. Comparison with Fig. 2 shows that, whereas a large number of sources ($\sim$800) are found with NIR excesses in the Carina Nebula field, the reference field shows sources to be concentrated mainly in the reddening band (the ``F'' region), with hardly any of them showing NIR excess.

\subsection{Color-Magnitude Diagram}

Figure 4 shows the color-magnitude diagram of the 17,495 sources detected in the $JHK_s$ bands with photometric errors smaller than 0.1 mag, sources selected from 2MASS, plus some 
1178 sources which are not detected in the $J$ band but are still detected in the $H$ and $K_s$ bands. The vertical solid lines (from left to right) in the figure represent the main-sequence tracks \citep{koornneef} reddened by $A_V$ equal to 0, 15, 30, and 45 mag for a distance of 2.5~kpc. The slanting lines
trace different spectral zones. The symbols in the plot mean the following: (a) known OB stars \citep{mj93} are shown as open squares, (b) the NIR counterparts of the X-ray sources are shown as pluses, (c) the Class II candidates selected from Fig. 2 are shown as crosses, (d) the Class I candidates selected from Fig. 2 are shown as diamonds, (e) the NIR sources with $H-K_s >2$, referred to as red sources from now onward, are shown as open circles, and (f) all other NIR sources are shown as dots.

The color-magnitude diagram appears to show four different groups. Most of the sources below spectral type B3, lying close to the main sequence with nearly zero extinction, are field stars forming the first group. The NIR sources with X-ray counterparts form the second group. As has been shown by, e.g., \citet{getman06}, the contamination due to field population is small in the X-ray samples of star-forming regions. Most of our NIR sources with X-ray counterparts should be largely predominated by the weak-lined T Tauri population.  
The Class II candidates selected based on the NIR colors, most of them with no X-ray counterpart, form the third group in the diagram, and these sources have extinction values of $A_V \gtrsim $ 7 mag. 
The fourth group is the red sources with $H-K_s>2$, and $A_V \gtrsim$ 30 mag. Table~2 lists the PMS candidates identified based on their NIR colors, i.e., Class II and Class I candidates, and based on the X-ray and NIR properties, i.e., Class III and Class II candidates.

\subsection{Interstellar Extinction}

We used the color-color diagram (Fig.~2a) to estimate the mean extinction toward the Carina Nebula. For this, we traced the stars in the reddening band to the solid line shown in the figure, near the main-sequence locus (M0-M6). The mean extinction determined from the above method is $A_V\sim$2.5 mag. The extinction toward the Carina Nebula has been known to be low in general despite the prominent nebulosity seen in the optical images. \citet{mj93} reported the mean $E(B-V)$ of the known OB stars to be $\sim$0.55, which assuming an $R=3.2$, gives $A_V\sim$1.5 mag. \citet{tapia03} found the mean extinction toward Tr\,14 and Tr\,16 to be $A_V\sim$2.5 mag. In our calculation, the mean extinctions estimated individually for Tr\,14 and for Tr\,16, show no significant difference. However, for the Tr\,16-SE group, which appears to be partially embedded, we found a higher mean extinction $A_V\sim$6 mag using the same method. As discussed in Paper I, the Tr\,16-SE group appears to be correlated with the southern molecular cloud, and ``sandwiched''
between two CO peaks. The individual extinction of stars in our field ranged from $A_V$ value of 0 to 40 mag.

\subsection{Spatial Distribution of Stellar Sources}

Figure 5 shows the mosaic image of the Carina Nebula in the $K_s$ band centered on $\eta$ Carinae (same as Fig.~1). The known OB stars together with the Class II and Class I candidates identified based on their NIR colors, are separately marked in the figure. Additionally marked are the red sources ($H-K_s>2$) as well as some 107 faint ($K_s > $17 mag) NIR sources.
The white contours represent the $^{12}$CO(1-0) emission \citep{brooks98}. The contours seen in the north-west and in the south-east trace parts of the northern and southern molecular clouds respectively, with Tr\,16 in between, where the molecular gas exists only in dense clumps of typical masses $\sim$10 \Msun~ \citep{cox95,brooks98}.

The known OB stars are seen to be distributed toward the clusters.
Most of the Class II and Class I candidates are also seen to be distributed from the north-west to the south-east across the image, connecting Tr\,14 and Tr\,16, with many more Class II candidates seen in Tr\,14 than in Tr\,16. However, the red sources with $H-K_s>2$ are concentrated to the
south-east of Tr\,16, near the denser part of the southern molecular
cloud, and near the three MSX point sources. The red sources to the south-east of Tr\,16, apart from showing a good correlation with the molecular cloud distribution, are also located at where the ionization
fronts are interacting with the cloud, as delineated by the PDRs 
seen in the MSX study \citep{smith00,rathborne02}. Moreover, in this
region an embedded IRAS source, IRAS 10430-5931, was found to be
associated with a bright rimmed globule \citep{megeath} along with
several bright X-ray sources, most of which are massive star candidates (Paper I)
with one known O4 star \citep{rgsmith}. Thus, whereas for some sources the large $H-K$ colors could be caused due to reddening by the cloud, many red
sources seen here should be PMS objects still embedded in clouds, representing the current generation of star formation.
The concentration of red sources near the MSX point sources 
G\,287.63$-$0.72, and G\,287.51$-$0.49 is particularly intriguing as these two MSX sources along with G\,287.47$-$0.54 are probably embedded clusters based on their mid-IR fluxes \citep{rathborne}.

The spatial distribution of the X-ray sources also presents a similar picture. Fig.~6 shows all the X-ray sources with/without NIR counterparts overlaid on the $K_s$ band image. The X-ray sources with NIR counterparts, most of which are low-mass PMS objects, are seen in the direction of Tr\,14 and Tr\,16. For the X-ray sources without NIR counterparts, their nature remains uncertain in general. It is, however, interesting to note the presence of some 5 such sources, one of which is a hard X-ray source, near G\,287.47$-$0.54 and G\,287.51$-$0.49. There are two more X-ray sources near these MSX sources, which are relatively bright and have NIR counterparts. All these 7 X-ray sources detected around the MSX sources are listed in Table~3 with their $Chandra$ counts and other properties. 

For G\,287.51$-$0.49, two of the X-ray sources in its vicinity are both hard X-ray sources with the hardness ratios, $HR$=0.17, and 0.28. The hardness ratio is defined as $HR$=$[C_{(2.5-8.0)}-C_{(0.5-2.5)}]/[C_{(2.5-8.0)}+C_{(0.5-2.5)}]$, where $C$ stands for the $Chandra$ counts in a given energy range shown in the subscript in keV. The high hardness ratios of these X-ray sources are indicative of very high absorbing environments or these sources have intrinsically hard spectra or both. The X-ray source No. 1 (see Table~3), which is relatively faint in X rays, with no NIR counterpart even in the $K_s$ band, is a candidate for Class I or Class 0 type object. The other hard X-ray source No. 2 is a bright X-ray source with an NIR counterpart and has a positional coincidence of 0.2$\arcsec$ with the MSX source G\,287.51$-$0.49 itself. For G\,287.47$-$0.54, we see four X-ray sources (Nos. 3 to 6), all being moderately bright in X rays and do not have NIR counterparts. Two of these have relatively high hardness ratios. For G\,287.63$-$0.72,  there is only one bright X-ray source nearby, No. 7, with an NIR counterpart.

Thus, the distribution of red NIR sources and X-ray sources strongly indicates that these MSX sources are embedded clusters and are sites of ongoing cluster formation. Apparently, we are seeing star formation at different evolutionary stages. Tr\,16 seems to be the most evolved as is evident from the large cavity it has created in the molecular cloud. Tr\,14 should be younger than Tr\,16 as is indicated by its association with its parental molecular cloud, and by the presence of a larger population of T Tauri candidates as compared to Tr\,16. Finally, the region south-east of Tr\,16 (including the Tr\,16-SE group) and near the MSX point sources show few T Tauri candidates but a concentration of an even younger generation of stellar population including red and faint NIR sources, and hard X-ray sources.

\subsection{Stellar Mass Estimates}

\subsubsection{Tr\,14}

We study the stellar populations of the clusters, Tr\,14 and Tr\,16 individually. To select the NIR sources associated with the clusters,
we have used the positions and the extents of the clusters 
from \citet{tapia03}, where they used the star counts in the $V$ band to determine the centers and the extents of these clusters. For Tr\,14, we used the
sources located within a 264$\arcsec$ radius of the cluster center, R.\,A.=$10^{h}43^{m}55.4^{s}$ and Decl.=$-59^{\arcdeg}32^{\arcmin}16^{\arcsec}$ (J2000). There are 3490 sources in this area centered on Tr\,14 with photometric errors smaller than 0.1 mag in all the $JHK_s$ bands. Figure~7 shows the NIR 2-color diagram of these sources. 
The colors of dwarfs and giants as well as the CTTS locus are plotted in a similar way as in Figs.~2 and 3.
There are 14 sources cross-identified with known O or B stars \citep{mj93}, shown as open squares, and 200 NIR sources with X-ray counterparts, marked as pluses.  
The T Tauri candidates based on the NIR colors in Tr\,14 region are selected in a similar way as discussed in \S{3.1}. We identified $\sim$219 CTTS candidates.

Figure~8 shows the color-magnitude diagram, $J$ versus $J-H$, for Tr\,14. The CTTS candidates selected from Fig.~7 are shown as crosses, the known OB stars are shown as open squares, and the NIR sources with an X-ray counterpart are shown as  pluses. 
We compare the distribution of sources with a 4 Myr post--main-sequence isochrone (dashed-dotted line) which was derived from \citet{bertelli} for a distance of 2.5~kpc, and 0.1 Myr (dotted line), 1 Myr (solid line) and 3 Myr (dashed line) PMS evolutionary tracks, which were derived using the model of \citet{palla99}. The PMS as well as post--main-sequence isochrones were corrected for extinction and reddening, by using the mean extinction value of $A_V$ of 2.5 mag estimated for this region, as discussed in \S{3.3}. The OB stars fit well the 4 Myr post--main-sequence isochrone. The low-mass PMS population, which is represented by the T Tauri candidates selected based on their NIR colors, and sources with X-ray counterparts, are seen to be distributed around 0.1--3 Myr PMS tracks. Masses range from 0.1 to 4 \Msun from bottom to top for all the PMS curves. For a representative age of $\sim$1 Myr,
more than $\sim$80 \% of the PMS candidates have masses smaller than 1 \Msun. The lowest mass limit of our data corresponding to our 10 $\sigma$ limiting magnitude in $J$, turns out to be $\sim$0.2 \Msun~ for a distance of 2.5~kpc of this cluster, assuming an age of 1 Myr.

\subsubsection{Tr\,16}

The NIR sources located within a 320$\arcsec$ radius of the cluster center, i.e., R.\,A.=$10^{h}~45^{m}~10.6^{s}$ and Decl.=$-59^{\arcdeg}~42^{\arcmin}~28^{\arcsec}$ (J2000) \citep{tapia03} were used to study Tr\,16. There are 4904 sources detected in this area centered on Tr\,16 in $JHK_s$, with photometric errors smaller than 0.1 mag. Figure~9 shows the NIR 2-color diagram for Tr\,16, with symbols the same as in Fig.~7. There are 
27 O or early-B stars and some 218 NIR sources with X-ray emission in Tr\,16. By using the similar criteria to identify the CTTS candidates as discussed in \S{3.1}, we selected 114 CTTS candidates.

Figure~10 shows the color-magnitude diagram, $J$ versus $J-H$, of Tr\,16. The symbols used in this figure are the same as in Fig.~8. In case of Tr\,16 also, the 4 Myr post--main-sequence track fits the massive stars, whereas the PMS population is distributed around 0.1--3 Myr PMS tracks. For Tr\,16, again using a representative age of $\sim$1 Myr, more than $\sim$77 \% of the PMS candidates have masses smaller than 1 \Msun. The comparison of Fig~10 with Fig.~8 shows a more noticeable background contamination in Tr\,16 than in Tr\,14, i.e., the scatter seen for $J-H > 2$. For Tr\,14, the molecular clouds apparently block a larger fraction of background stars.

\subsection{The $K_s$-band Luminosity Function}

The $K_s$-band luminosity function (KLF) of an embedded cluster is useful in constraining the age and initial mass function (IMF) of the cluster. In order to convert the observed KLF
of a cluster --- which is the luminosity function of all the stars detected in the $K_s$ band in the direction of the cluster --- to the cluster KLF, one needs to apply corrections for (a) the incompleteness of the star counts as a function of the $K_s$ magnitude, and (b) the field star contribution in the line of sight of the cluster.  

The completeness of our $K_s$-band data was estimated for each individual field using the ADDSTAR package in IRAF, as described in \S{2.1.2}. The completeness of fields, Nos. 1, 3, and 5 (see Table~1), are demonstrated in Fig.~11, from which one sees a systematic lower completeness for the relatively crowded field No. 3, which includes Tr\,14, than that of either field No. 5 (including Tr\,16) or field No. 1. This completeness correction estimated for the clusters was applied to the observed KLFs of each cluster. 

In the studies of the luminosity function of a star cluster, the correction for the foreground and background population in the line of sight of the cluster is of crucial importance. The under/over estimate of the field stars in the line of sight of a particular cluster can significantly change the estimated cluster luminosity function from the observed one. One method to correct for the foreground and background contamination is to use a reference field, located reasonably away from the cluster, and yet close enough to have a similar Galactic field star distribution, e.g., with the same Galactic latitude. This reference field is then assumed to represent the field population to the cluster. However, a young and partly embedded cluster is still associated with the parental molecular cloud and thus renders an appreciable, and often variable, amount of extinction for the background field stars.

To correct for the foreground and background contamination, we made use of the 
Galactic model by \citet{robin03}.
This Galactic model makes use of the Besan\c{c}on model of population synthesis to reproduce the stellar contents of the Galaxy. The advantage in using this model is that the background stars (d $>$ 2.5 kpc) can be separated from the foreground stars (d $<$ 2.5 kpc). While all the stars in the field suffer a general interstellar extinction, only the background stars suffer an additional extinction due to the cloud --- the Carina molecular cloud complex in our case.
The use of the model allows us to apply an additional cloud extinction to the background stars. To ensure that the model does a reasonably good job in predicting the field star population, we compared the model star counts with our observed reference field.
Figure~12 shows the comparison of the completeness corrected KLF of the reference field with (a) the model star counts where no extinction has been applied to the model, and (b)
where all the model stars are made fainter by an interstellar extinction of $A_V$ = 1 mag~kpc$^{-1}$. It can be seen that the latter matches reasonably well with the observed reference field. Hence, we used this model to predict the contamination to the Carina Nebula field, for which all the model stars are made fainter by a general interstellar extinction of $A_V$ = 1 mag~kpc$^{-1}$, whereas the background stars are made fainter by an additional cloud extinction. 

To estimate the cloud extinction which affects the background population in the direction of the clusters, we used the NICE (Near-Infrared Color Excess) method \citep{lada92}, in which the excess in the mean $H-K$ color of the target fields relative to that of the reference field is used to estimate the extinction due to the cloud. Using this method, we found a moderate $A_V \sim$1.5 mag for Tr\,16, a higher value of $A_V \sim$2.5 mag for Tr\,14 and $A_V \sim$5.5 mag for the Tr\,16-SE group.
It should be noted that the mean extinction estimated for the individual clusters and for the Tr\,16-SE group in \S{3.3} represents the mean extinction suffered by the cluster members, whereas the NICE method  
gives an estimate of the cloud extinction which affects the background stars. While a moderate extinction of $A_V \sim$1.5 mag (or $A_K \sim$0.15 mag) for Tr\,16 or 
$A_V \sim$2.5 mag for Tr\,14 may not have significant effects on the field star correction, the additional cloud extinction in case of the Tr\,16-SE group significantly alters the shape of the cluster KLF.

In Figure~13a, the observed KLF of Tr\,16 is shown with only completeness corrected KLF, and our estimate of the field star contribution. Fig.~13b shows the cluster KLF, where the field star contribution has been subtracted from the completeness corrected KLF of Tr\,16. The cluster KLF is seen to rise smoothly until $K_s \sim$16.5 mag and turns over at $K_s$ = 16.5--17 mag. This turnover is likely due to the sensitivity limit of our data and not the intrinsic turnover for the cluster. Our completeness for the Carina Nebula field reaches 70\% at $K_s \sim$16.75 mag. For comparison, the completeness of 70\% for the reference field is reached at $K_s \sim$17.75 mag, and the reference field KLF is seen not to turn over until $K_s$ = 18 mag (see Fig.~12). Thus, it appears that the turnover of Tr\,16 KLF is due to the sensitivity of our data.

The observed KLF of Tr\,14 is shown in Fig.~14a, together with only 
completeness corrected KLF and the field star contribution. The cluster KLF is shown in Fig.~14b. One can readily see that the Tr\,14 KLF has a very different shape from the Tr\,16 KLF (Fig.~13b). For Tr\,14, the KLF rises up to $K_s \sim$14.5 mag and then levels off until a sharp peak at $K_s$ = 16--16.5 mag, after which it turns over. The peak stands out even in the observed KLF (Fig.~14a), where some 1365 stars are detected in this magnitude bin (16--16.5), whereas after the completeness correction the number becomes 1706. No matter whether we subtract either the number of stars in the observed reference field or the field star contribution from the model, there is an excess of some 1000 stars in this bin. \citet{zinnecker93} found sharp peaks in the synthetic KLFs of extremely young (less than 1--2 Myr) clusters, which the authors attributed to the onset of deuterium burning. They found the deuterium peak to shift to fainter absolute magnitudes as the cluster ages, and disappear long before the cluster reaches an age of 10 Myr. We suggest the peak seen in the Tr\,14 KLF to be the deuterium peak as has been evidenced in other star clusters of ages 1--2 Myr e.g., in \citet{muench03}. The turnover of the Tr\,14 KLF occurs at a slightly brighter magnitude than the Tr\,16 KLF, which, given the similar completeness limits of our data for the two clusters, implies a younger age for Tr\,14 than for Tr\,16.

Finally, the observed KLF of the Tr\,16-SE group is shown together with only completeness corrected KLF and the field star contribution in Fig.~15a. The KLF of the Tr\,16-SE group is shown in Fig.~15b. We see a clear excess of stars in the Tr\,16-SE group from the bright end to the faint end as compared to the field stars, which suggests that the group should indeed be an embedded association. The KLF is seen to rise until $K_s$ = 16 mag, at which it seems to turn over, but to rise again at $K_s$ = 16.5--17 mag. The turning over at $K_s$ = 16 mag can also be seen in the raw KLF, where a small dip is seen at $K_s$=16--16.5 mag. The rise seen at $K_s$ = 16.5--17 mag after the apparent turnover at $K_s$ = 16 mag might imply a non-coeval formation for this embedded group.

The KLFs of young embedded clusters are known to follow power-law shapes \citep{elada91,clada93}. Fig.~16 shows the cluster KLFs of Tr\,16 and Tr\,14, and of the Tr\,16-SE group. A power-law with an index $\alpha$ has been fitted to each KLF, where $\alpha$ is defined as, 
$dN(m_K)/dm_K\propto10^{\alpha m_K}$, with $N(m_K)$ being the number of stars brighter than $m_K$. The KLF index $\alpha$  is found to be 0.33 $\pm$ 0.01 for Tr\,16, and slightly steeper for Tr\,14, $\alpha$ = 0.37 $\pm$ 0.01. For the Tr\,16-SE group, KLF slope is found to be $\alpha$ = 0.30
$\pm$ 0.03. These values of KLF slopes are in agreement with $\alpha$=0.32--0.40,
the values reported for other young embedded stellar clusters, such as NGC~2024, NGC~2068, and NGC~2071 \citep{elada91}, and NGC~2264 \citep{clada93}.

Using the KLFs of the clusters Tr\,16 and Tr\,14, and of the Tr\,16-SE group,
we constructed their approximate mass functions assuming certain mass-luminosity relations. Given the young ages of these clusters, these mass functions correspond to the IMFs of these clusters. We estimated the stellar masses from the $K_s$ magnitudes 
by using the PMS isochrone from \citet {palla99} for masses between 0.1 to 4 \Msun, and the post--main-sequence isochrone from \citet {lejeune01} for the higher mass stars. Assuming a 
representative age of 3 Myr for the clusters, and a distance of 2.5~kpc, the resulting mass functions of Tr\,16, Tr\,14, and the Tr\,16-SE group, are shown in Fig.~17.
The dashed line in the figure corresponds to the 90\% completeness limit of the data, $\sim$0.7 \Msun. The best linear fit gives a slope of $-$1.62 $\pm$ 0.13 for Tr\,16, $-$1.69 $\pm$ 0.08 for Tr\,14, and $-$1.58 $\pm$ 0.12 for the Tr\,16-SE group. If we consider an age of 1 Myr instead, the derived slopes of the mass functions become $-$1.34 $\pm$ 0.13 for Tr\,16, $-$1.39 $\pm$ 0.08 for Tr\,14, and $-$1.30 $\pm$ 0.12 for the Tr\,16-SE group. Given a KLF, the derived mass function of a cluster has a shallower slope for a younger age, which can be understood as PMS stars in general get fainter as they evolve toward the main sequence, and more so for lower-mass stars. Considering all the assumptions and uncertainties involved in deriving these mass functions, we conclude that these slopes are in agreement with the slope for field star IMF, $-1.35$, derived by \citet{salpeter55}.


\subsection{Individual Sources and Regions}

We discuss below a few infrared sources in our field, i.e., IRAS\,10430-5931, and the MSX sources, G\,287.51-0.49 and G\,287.47-0.54, where we find indications of ongoing cluster formation. The MSX sources are found to be candidate embedded clusters based on their mid-infrared fluxes \citep{rathborne}. In our analysis we find concentration of red NIR sources and X-ray sources around these MSX sources. We present the $K_s$ band images around these sources to show the distribution of the young star candidates we identified. Other than the MSX and IRAS sources, we also discuss the Tr\,16-SE group in detail, which was first noticed in our earlier work (Paper I) as an embedded young stellar group or cluster.

\subsubsection{IRAS\,10430-5931}

IRAS\,10430-5931 (R.\,A=$10^h45^m36^s$, Decl.=$-59^{\arcdeg}47^{\arcmin}02^{\arcsec}$ (J2000)) is a deeply embedded infrared source, which is associated
with a bright rimmed globule \citep{megeath}. The flux of this IRAS source increases from 12 $\mu$m to 100 $\mu$m, which also suggests
the source to be an embedded young star. From the molecular line observations $^{12}$CO(1--0), $^{13}$CO(1--0), and $^{18}$CO(1--0), \citet{megeath} estimated an average extinction of $A_V$=15 mag for this globule and $A_V$=25 mag near the center of this globule.
They found the size of the globule to be $\sim$0.5 pc and derived its mass to be $\sim$67 
\Msun~ using the $^{13}$CO data. This IRAS source (Fig.~18a) is associated with the MSX point source G\,287.63-0.72, located about 10$\arcsec$ away \citep{rathborne}. 
The bright rim is clearly visible and a number of unresolved $K_s$ sources are seen distributed near the head of the rim. Two T Tauri candidates are distributed outside the rim, which might represent the recent star formation, whereas 
a handful of red NIR sources ($H-K_s > 2$), the MSX source, as well as the IRAS source are located inside the bright rim, which may represent the present epoch of star formation.

\subsubsection{G\,287.51-0.49}

G\,287.51-0.49 is an MSX point source which is a candidate embedded cluster based on its mid-IR fluxes \citep{rathborne}. Fig. 18b shows the $K_s$-band image centered on this MSX point source. The two hard X-ray sources, as well as the T Tauri candidates and the red NIR sources identified around this MSX source are marked in the figure. It appears that this group of young stars that we have identified, is part of an early episode of cluster formation.

\subsubsection{G\,287.47-0.54}

In Fig. 18c the $K_s$ band image around the MSX point source, G\,287.47-0.54, is shown. This mid-IR source also has mid-IR fluxes consistent with being an embedded cluster \citep{rathborne}. From a SIMBAD search, we found an IRAS source, IRAS\,10424-5916, located about 40$\arcsec$ away, in the north-west of the MSX source. This infrared source has IRAS colors consistent with that of an embedded star. There are four X-ray sources, none with an NIR counterpart, some T Tauri candidates and red NIR sources in the vicinity of this MSX source. Prominent diffuse NIR emission can be seen surrounding this MSX source, enveloping many unresolved $K_s$ sources.

\subsubsection{Tr\,16-SE group}

The $K_s$ band image around a compact group of ten X-ray sources (Paper I) is shown in Fig. 18d. One of these X-ray sources is a known O4 star (Paper I) with the rest mostly being candidate massive stars. The NIR 
sources with X-ray emission are marked in the figure along with the T Tauri candidates and the red sources. The KLF of this group shows a clear excess of sources as compared to the field stars, suggesting that it should be a bona fide young star group or cluster.
The KLF slope of this group, $\alpha$= 0.30, is indicative of its young age.

\section{Summary}

An NIR imaging survey of a large $\sim$400 arcmin$^{2}$ area centered on $\eta$ Carinae, including clusters Tr\,14 and Tr\,16 is presented. The 10 $\sigma$
limiting magnitudes of the data are $\sim$18.5, 17.5, and 16.5 mag, in the $J$, $H$, and $K_s$ bands, respectively. The NIR observations presented in this work are
deeper than any NIR imaging survey to date, for this large field in the Carina Nebula. The main conclusions derived from this work are the following.

\begin{enumerate}
\item Using the NIR color diagnostics, we found 544 Class II candidates and 11 class I candidates in the entire observed field. A large number ($\sim$40) of very red NIR sources with $H-K_s > 2$, are found, apart from some 100 faint NIR sources ($K_s > 17$) which are detected only in the $K_s$ band.
\item The spatial distribution of the PMS candidates, red and faint NIR sources, suggests that we are seeing star formation in different evolutionary 
stages in this region. The T Tauri candidates are seen to be distributed in the direction of the clusters, with Tr\,14 containing a much larger population of Class II candidates than Tr\,16 does. Most of the red and faint NIR sources are seen to the south-east of Tr\,16. The red sources together with some hard X-ray sources are also found near the three MSX point sources, hinting that these mid-infrared sources are sites of ongoing cluster formation.
\item The massive stars of the clusters can be fitted with a 4 Myr post--main-sequence isochrone of \citet{bertelli}, whereas the PMS population is seen to be distributed along 0.1--3 Myr PMS isochrones \citet{palla99}. 
\item The shapes of the KLFs of Tr\,16 and of Tr\,14 differ significantly. The Tr\,14 KLF shows a sharp peak in the $K_s$=16--16.5 mag likely due to the deuterium burning PMS stars, just before it turns over. For Tr\,16, the KLF rises smoothly before it turns over at $K_s \sim$17 mag, likely due to the sensitivity of our data. 
\item The KLF of the Tr\,16-SE group shows that the group has a much higher number density of $K_s$ sources relative to the field therefore should be a bona fide young star group or cluster. The KLF slope for the Tr\,16-SE group is found to be 0.30, which is indicative of its very young age.
\item The slopes of the IMFs are found to be $-$1.62 for Tr\,16, $-$1.69 for Tr\,14, and $-$1.58 for the Tr\,16-SE group, down to our completeness limit of $\sim$0.7 \Msun.
\end{enumerate}

\acknowledgments 
In this work, we have made use of the $Chandra$ archival data. We thank Kate Brooks for providing us with the
$\mathrm{~^{12}CO(1-0)}$ data of the Carina Nebula, which was obtained with the Mopra Antenna, operated by the Australia Telescope National Facility,
CSIRO during 1996--1997. We thank Annie Robin for letting us use her model of stellar population synthesis. We acknowledge Francesco Palla and Steven Stahler for the PMS model 
tracks and stellar isochrones. We thank the SAAO staff for maintenance of the IRSF
telescope. We thank the anonymous referee for valuable comments and suggestions.
KS would like to thank the Tata Institute of Fundamental Research for the kind hospitality during her visit at the institute, where a part of this work was carried out. KS and WPC acknowledge the financial support of the grants NSC95-2119-M-008-028 and NSC95-2745-M-008-002 of the National Science Council of Taiwan.

\clearpage

\begin{deluxetable}{lccccc}
\tablecaption{Log of Observations in the Carina Nebula}
\tabletypesize{\normalsize}
\tablenum{1}
\tablewidth{0pt}
\tablehead{
\colhead{Date} & \colhead{Field} & \colhead{$\alpha$} & \colhead{$\delta$} & \colhead{Seeing (FWHM)} & \colhead{Airmass}  \\ 

\colhead{$~$} &\colhead{$~$} & \colhead{(J2000)} & \colhead{(J2000)} &\colhead{\arcsec} & \colhead{$~$}}

\startdata

2003 April 1& 1...... & 10 45 52.80 & -59 33 56.4 & 1.5 & 1.2 \\
2003 April 1& 2...... & 10 45 00.76 & -59 32 06.6 & 1.5 & 1.2  \\
2003 April 2 & 3...... & 10 44 08.20 & -59 32 55.8 & 1.1 & 1.3 \\
2003 April 2 & 4...... & 10 45 46.13 & -59 39 23.0 & 1.1 & 1.3 \\
2005 January 14& 5...... & 10 45 04.05 & -59 38 58.4 & 1.1 & 1.1 \\
2005 January 14 & 6...... & 10 44 10.50 & -59 39 24.9 & 1.1 & 1.1 \\
2003 April 10& 7...... & 10 45 59.67 & -59 47 46.3 & 1.3 & 1.1 \\
2003 April 10& 8...... & 10 45 03.66 & -59 48 06.5 & 1.3 & 1.2 \\ 
2003 April 1&9...... & 10 44 05.89 & -59 46 36.0 & 1.4 & 1.3 \\
2005 January 14& Reference field &  10 47 17.41 & -59 39 26.0 & 1.1 & 
1.1 \\

\enddata

\tablecomments{Units of right ascension are hours, minutes, and seconds, 
and
units of declination are degrees, arcminutes, and arcseconds.}
 
\end{deluxetable}

\clearpage

\begin{deluxetable}{cccccccc}
\tablecaption{PMS candidates in the Carina Nebula\tablenotemark{a}}

\tabletypesize{\small}
\tablenum{2}
\tablewidth{0pt}
\tablehead{
\colhead{Number} &  \colhead{$\alpha$} & \colhead{$\delta$} & \colhead{$J$} &
\colhead{$H$} & \colhead{$K_s$} & \colhead{$Chandra$} & \colhead{$Chandra$} \\

\colhead{$~$} & \colhead{(J2000)} & \colhead{(J2000)} &\colhead{(mag)} &
\colhead{(mag)} & \colhead{(mag)} & \colhead{(counts)} & 
\colhead{(counts s$^{-1}$)}}

\startdata

1& 10 43 35.52& -59 38 12.4& 15.79& 14.43& 13.55& -& -\\
2& 10 43 35.71& -59 30 13.0& 16.56& 15.73& 15.23& -& -\\
3& 10 43 35.76& -59 34 14.1& 15.01& 14.03& 13.70& 61.6& 0.00103\\
4& 10 43 35.85& -59 34 44.8& 16.52& 15.30& 14.37& -& -\\
5& 10 43 35.92& -59 38 15.9& 17.23& 16.16& 15.39& -& -\\
6& 10 43 35.99& -59 39 04.6& 17.65& 15.83& 14.64& -& -\\
7& 10 43 36.28& -59 35 42.1& 15.45& 13.91& 12.96& -& -\\
8& 10 43 36.37& -59 31 40.5& 16.43& 15.32& 14.60& -& -\\
9& 10 43 36.46& -59 38 42.0& 17.21& 16.24& 15.49& -& -\\
10& 10 43 36.48& -59 31 59.0& 15.93& 14.94& 14.62& 42.3& 0.00071\\
11& 10 43 36.48& -59 33 42.3& 13.60& 13.18& 13.10& 28.4& 0.00047\\
12& 10 43 36.49& -59 34 35.4& 16.69& 15.78& 15.18& -& -\\
13& 10 43 36.60& -59 36 01.7& 17.18& 15.96& 15.20& -& -\\
14& 10 43 36.75& -59 38 26.3& 16.62& 15.54& 14.89& -& -\\
15& 10 43 36.84& -59 32 47.7& 12.73& 11.76& 10.90& 64.4& 0.00107\\
16& 10 43 36.84& -59 33 14.7& 12.56& 12.07& 11.90& 68.9& 0.00115\\
17& 10 43 36.84& -59 33 58.2& 16.49& 15.33& 14.94& -& -\\
18& 10 43 36.84& -59 34 37.4& 14.47& 13.45& 12.69& 50.3& 0.00084\\
19& 10 43 36.84& -59 36 08.2& 15.40& 14.08& 13.54& 86.1& 0.00144\\
20& 10 43 37.09& -59 45 06.9& 18.13& 16.43& 15.00& -& -\\
21& 10 43 37.17& -59 33 35.8& 14.96& 13.80& 13.00& -& -\\
22& 10 43 37.20& -59 32 30.4& 16.46& 15.47& 14.85& -& -\\
23& 10 43 37.20& -59 36 00.1& 17.77& 16.51& 15.73& -& -\\
24& 10 43 37.20& -59 32 11.7& 15.96& 15.14& 14.86& 30.5& 0.00051\\
25& 10 43 03.72& -59 34 24.4& 12.72& 12.32& 12.24& 43.0& 0.00072\\

\enddata

\tablenotetext{a}{This is only a partial list, the complete table is 
available at 
http://cepheus.astro.ncu.edu.tw/$\sim$kaushar/download/table2.pdf}

\tablecomments{Units of right ascension are hours, minutes, and seconds,
and units of declination are degrees, arcminutes, and arcseconds.}

\end{deluxetable}

\clearpage

\begin{deluxetable}{ccccccccc}
\tablecaption{The X-ray sources detected near the MSX point sources.}
\tablenum{3}
\tablewidth{0pt}
\tablehead{
\colhead{ID} & \colhead{$\alpha$} & \colhead{$\delta$} & \colhead{$Chandra$} & 
\colhead{$Chandra$} & \colhead{$HR$} & \colhead{$J$} & \colhead{$H$} & 
\colhead{$K_s$} \\

\colhead{$ ~$} & \colhead{(J2000)} & \colhead{(J2000)} &
\colhead{(counts)} & \colhead{(counts s$^{-1}$)} & \colhead{~} & 
\colhead{(mag)} & \colhead{(mag)} & 
\colhead{(mag)} }
\startdata
1 & 10 44 57.39 & -59 31 19.7 & 97.0 & 0.0016 & 0.28 & - & - & - \\
2 & 10 44 59.64 & -59 31 19.8 & 288.0 & 0.0048 & 0.17 & 12.5 & 11.5  & 
10.8 \\
3 & 10 44 32.39 & -59 33 01.5 & 117.0 & 0.0019 & -0.56 & - & - & -  \\
4 & 10 44 32.39 & -59 33 13.2 & 70.0 & 0.0011 & -0.08 & - & - & -  \\
5 & 10 44 33.15 & -59 33 27.7 & 71.3 & 0.0011 & -0.33 & - & - & -  \\
6 & 10 44 35.08 & -59 33 30.7 & 79.1 & 0.0013 & -0.91 & - & - & -  \\
7 & 10 44 57.94 & -59 47 09.4 & 164.5 & 0.0091 & -0.25 & 13.4 & 12.7 & 
12.3 \\
\enddata

\tablecomments{The $Chandra$ counts listed here are without background 
subtraction. $HR$ is the hardness ratio of the X-ray sources, 
defined as
$HR$=$[C_{(2.5-8.0)}-C_{(0.5-2.5)}]/[C_{(2.5-8.0)}+C_{(0.5-2.5)}]$, where $C$ 
stands for the $Chandra$ counts in a given energy range shown as the subscript
in keV.}
\end{deluxetable}

\clearpage

\begin{figure}
\begin{center}
\includegraphics[scale=0.8]{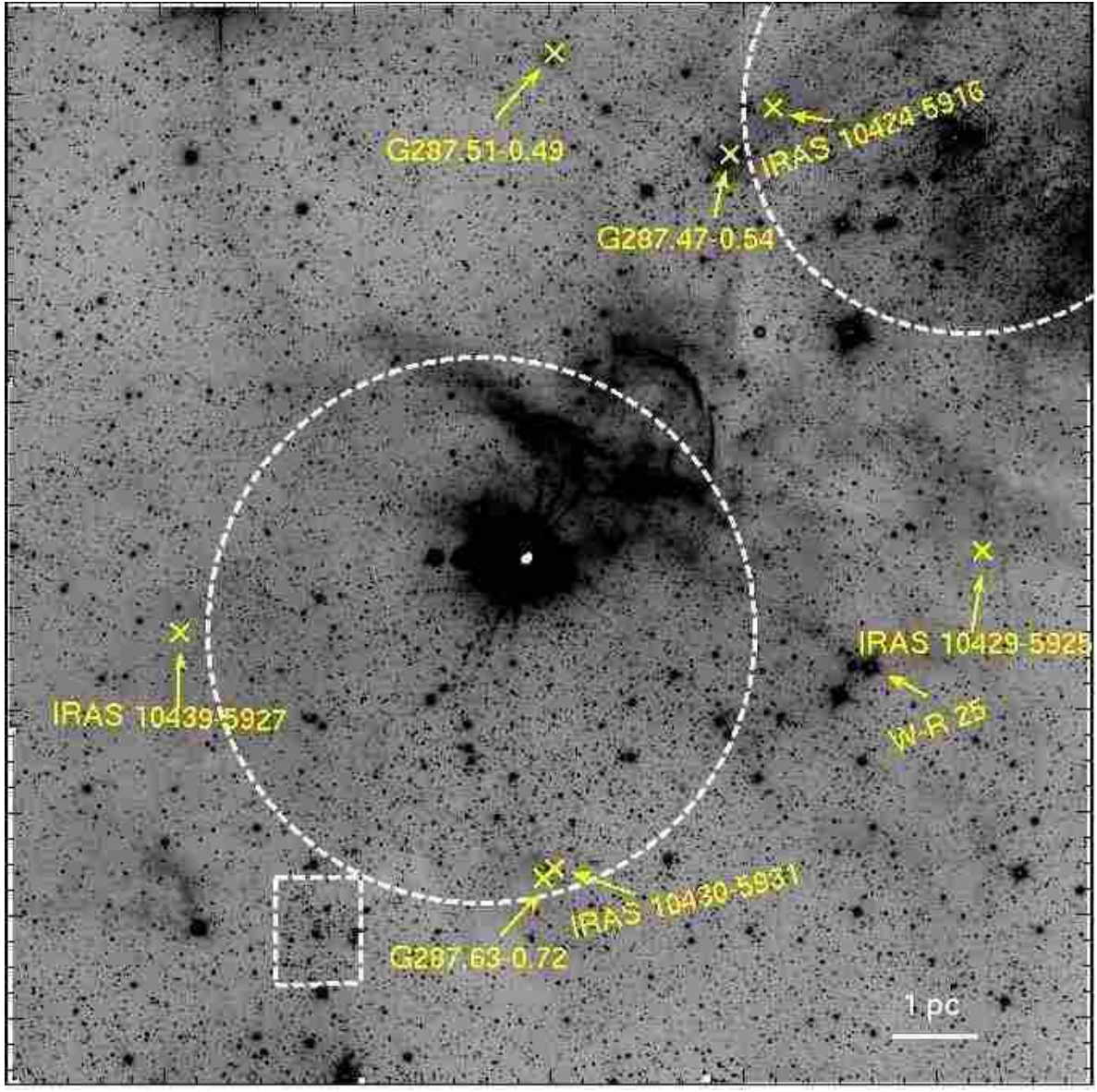}
\label{chandra}
\caption{The $K_s$ band mosaic image of the Carina Nebula. 
The image is centered on $\eta$ Carinae and has a field of view of 
$\sim$20\am $\times$ 20\am. 
The dashed circles in the center and in the top mark the 
extents of the clusters, Tr\,16 and Tr\,14, respectively in our study. The dashed rectangle 
marks the extent of the Tr\,16-SE group (see text). 
The locations of three MSX point sources, which are likely embedded clusters, 
and four IRAS sources in the field, are marked with crosses and labeled. 
The WR star, HD 93265 (WR 25), is marked as well. 
The MSX point source G287.63-0.72 is coincident with the IRAS source, IRAS 10430-5931.} 
\end{center}
\end{figure}


\clearpage

\begin{figure}
\begin{center}
\plotone{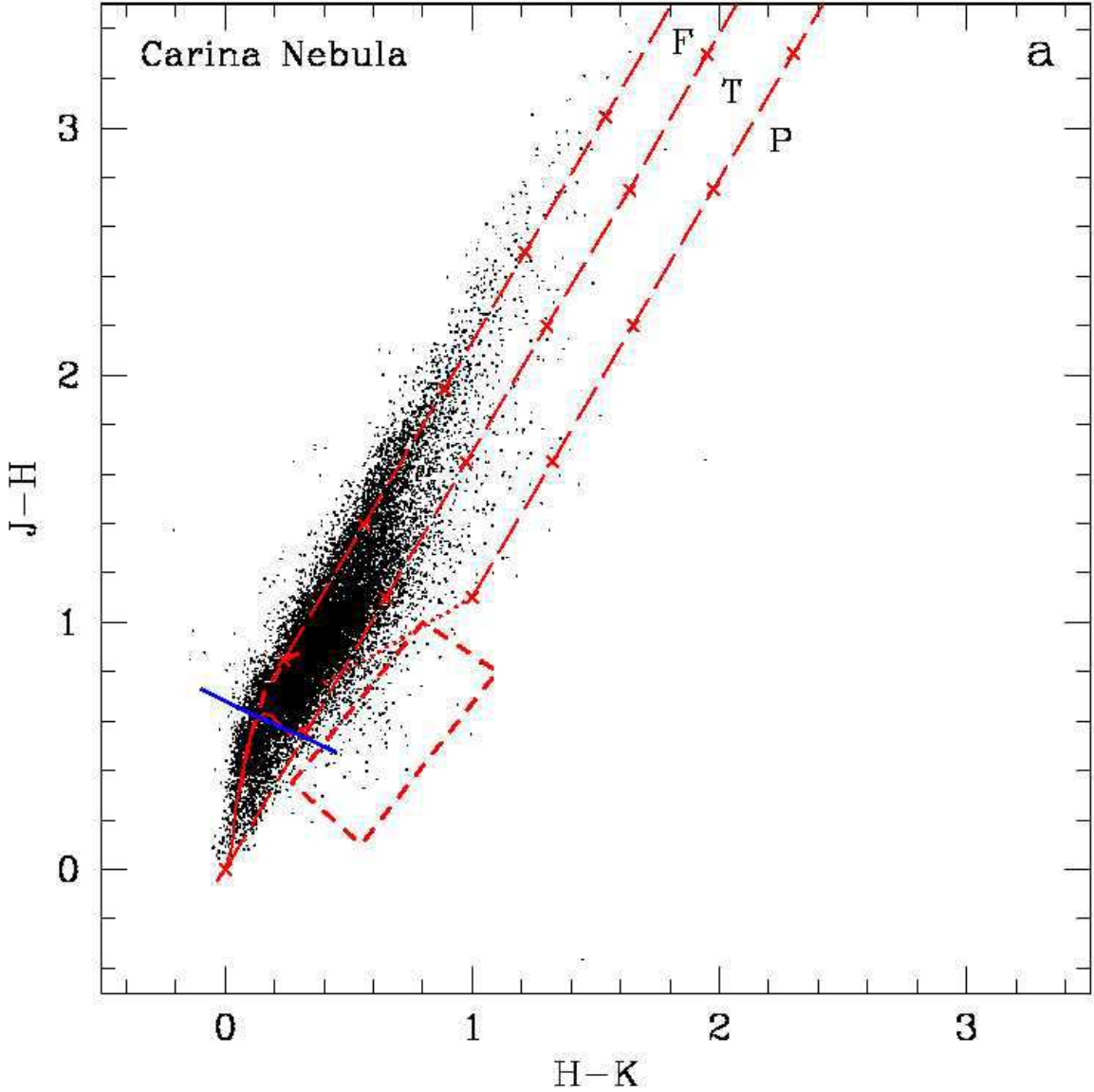}

\caption{The color-color diagram of (a) all the NIR sources in the composite field  with photometric errors less than 0.1 mag in all the $JHK_s$ bands and of the bright sources selected from 2MASS, (b) the OB stars in our sample, and (c) the NIR sources with X-ray counterparts. The dwarf and giant loci are shown as the solid and dashed curve respectively and are taken from \citet{bb} 
after converting to the CIT system. The long dashed lines represent the reddening vectors with the crosses marked on them separated by $A_V$ = 5 mag. In (a), and (c), the dotted line represents the locus of unreddened classical T Tauri stars \citep{meyer} and the region bounded by short dashed lines is where unreddened Herbig Ae/Be stars are found \citep{hernandez}. In (a), the plot is classified into three regions, ``F'', ``T'', and ``P'' (see text for details) and the solid line passing through the turn-off point of the main sequence locus is used for the estimate of extinction.}
\end{center}
\end{figure}

\clearpage

\begin{figure}
\setcounter{figure}{1}
\begin{center}
\plotone{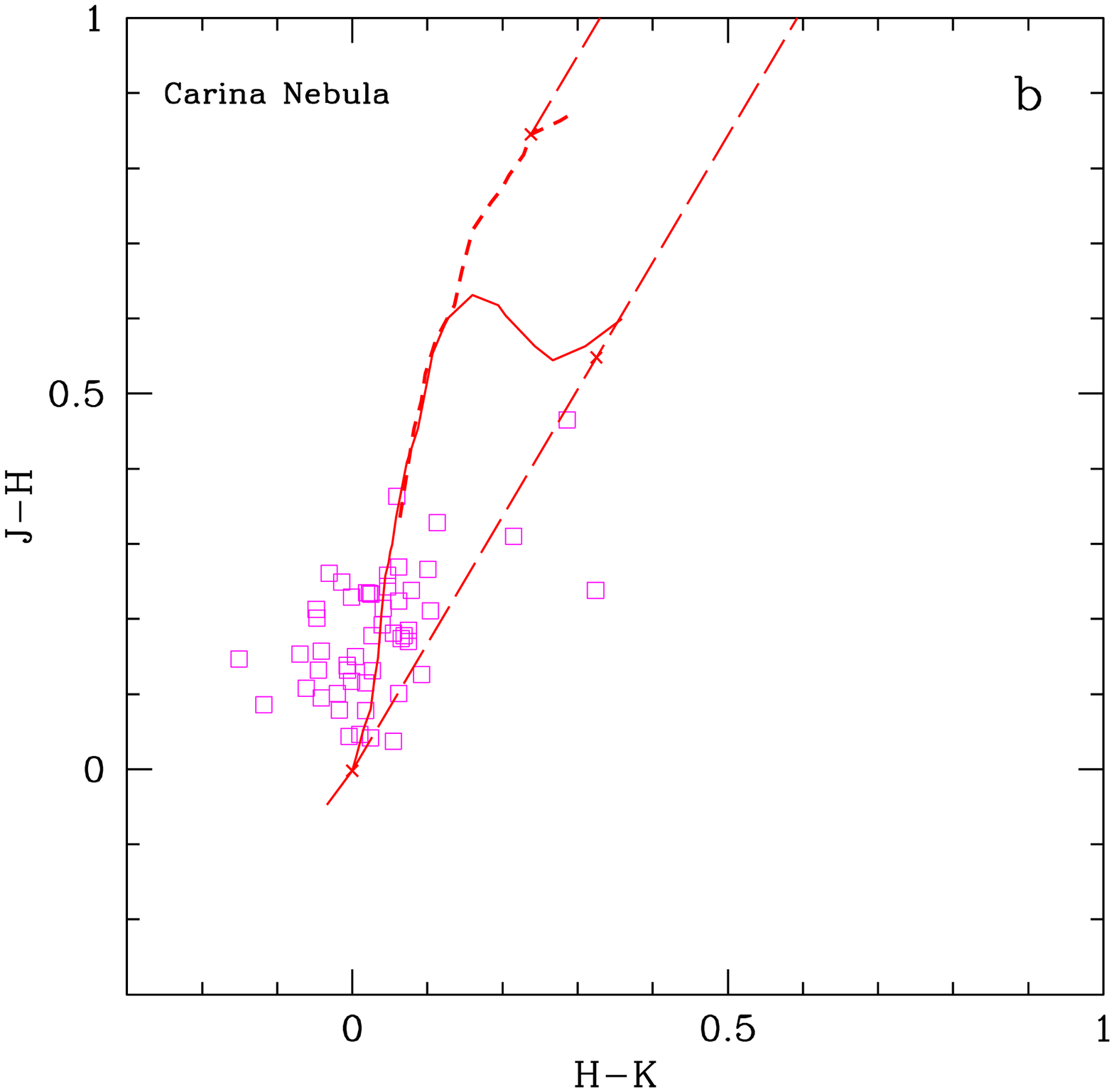}
\caption{Continued}
\end{center}
\end{figure}

\clearpage

\begin{figure}
\setcounter{figure}{1}
\begin{center}
\plotone{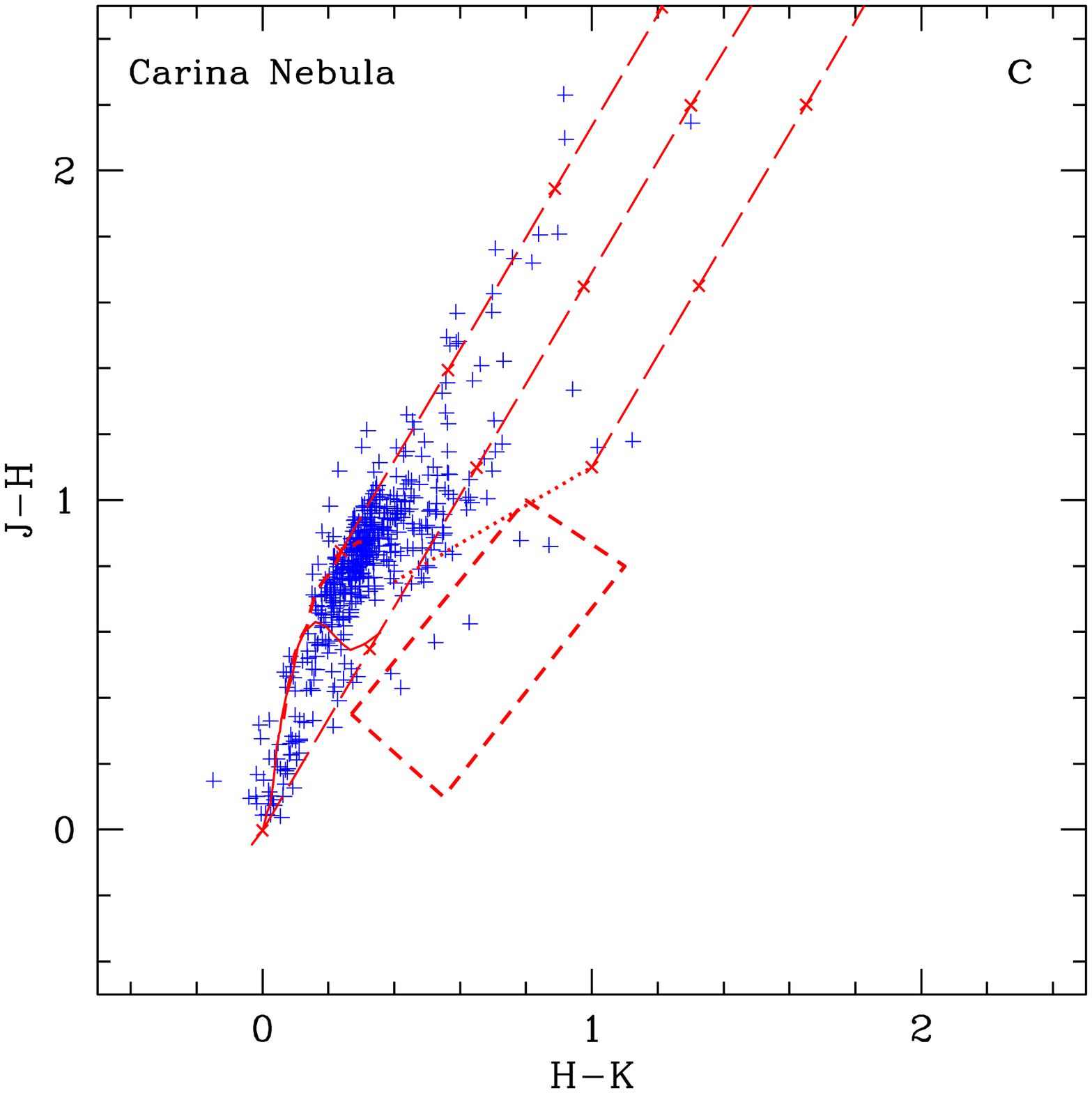}
\caption{Continued}
\end{center}
\end{figure}

\clearpage

\begin{figure}
\setcounter{figure}{2}
\begin{center}
\plotone{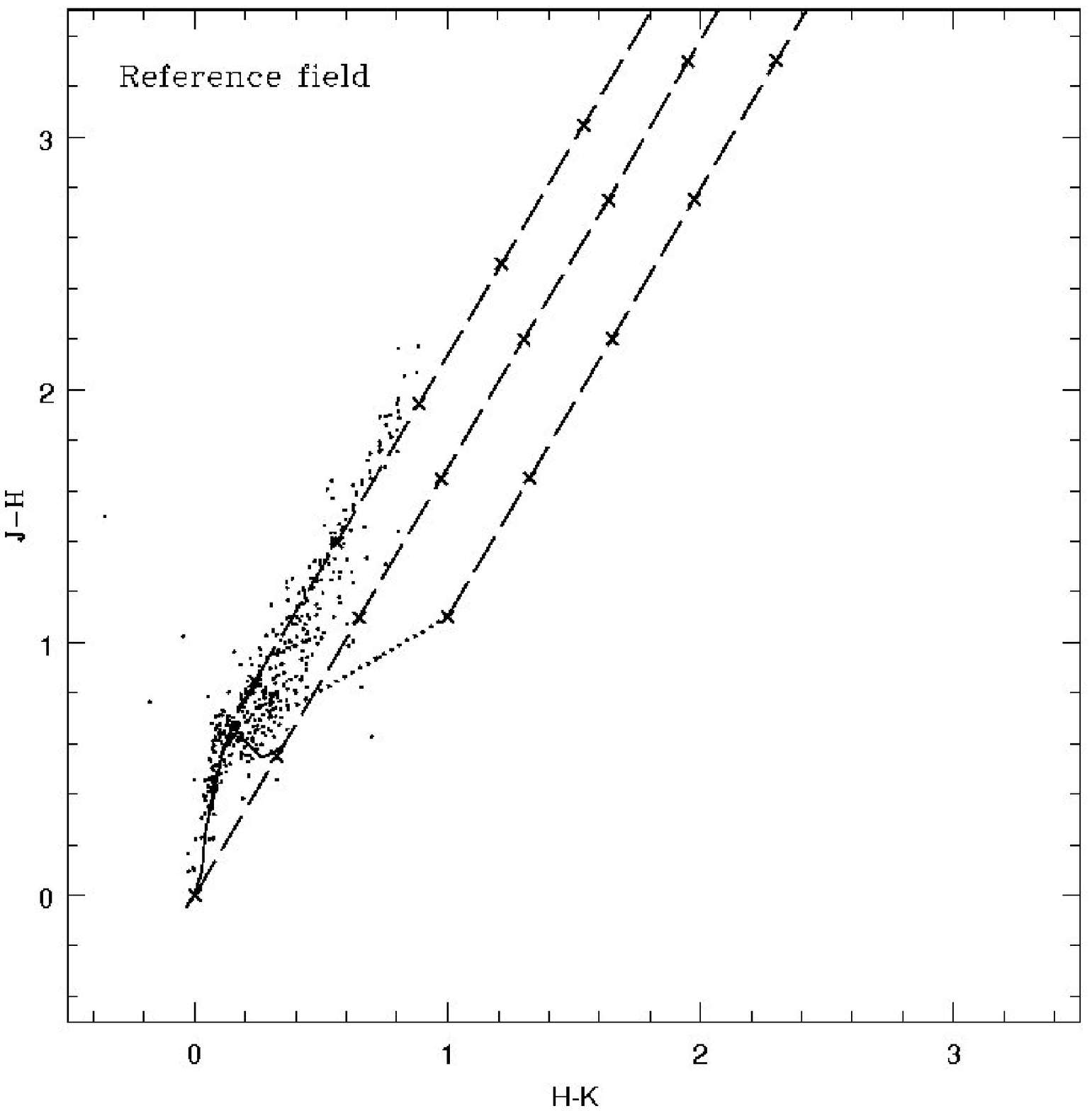}
\label{chandra}
\caption{The color-color diagram of the reference field (see text). The dwarf and giant loci, the reddening vectors and the classical T Tauri locus are plotted in a similar manner as in Fig.~2.
}
\end{center}
\end{figure}

\clearpage

\begin{figure}
\begin{center}
\plotone{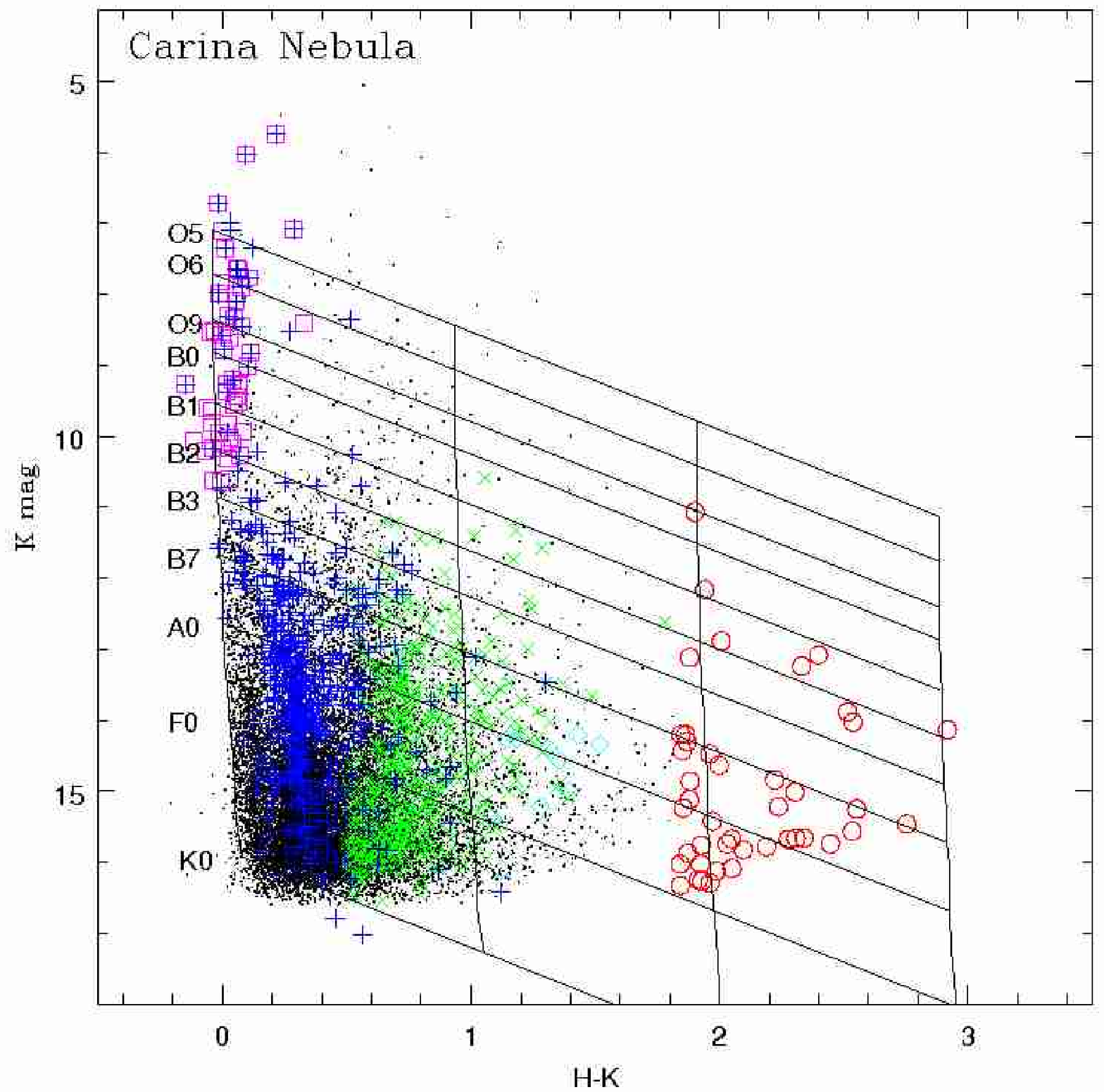}
\label{chandra}
\caption{The color-magnitude diagram of all the sources detected in $JHK_s$ with photometric errors
smaller than 0.1 mag, together with the sources detected only in $H$ and $K_s$, plus bright sources selected from 2MASS. The nearly vertical solid lines (from left to right) 
represent the main-sequence tracks \citep{koornneef} reddened by $A_V$ equal to 0, 15, 30 and 
45 mag for a distance of 2.5~kpc. Class II candidates selected from the 
color-color plot are shown as crosses, Class I candidates are shown as diamonds. Sources with $H-K_s>2.0$ (or $H-K>1.8$) are shown as circles. The known massive stars of O or B types are shown as squares, and the sources with X-ray counterparts are shown as pluses. All the rest of the NIR sources are shown as dots.
} 
\end{center}
\end{figure}

\clearpage

\begin{figure}
\begin{center}
\plotone{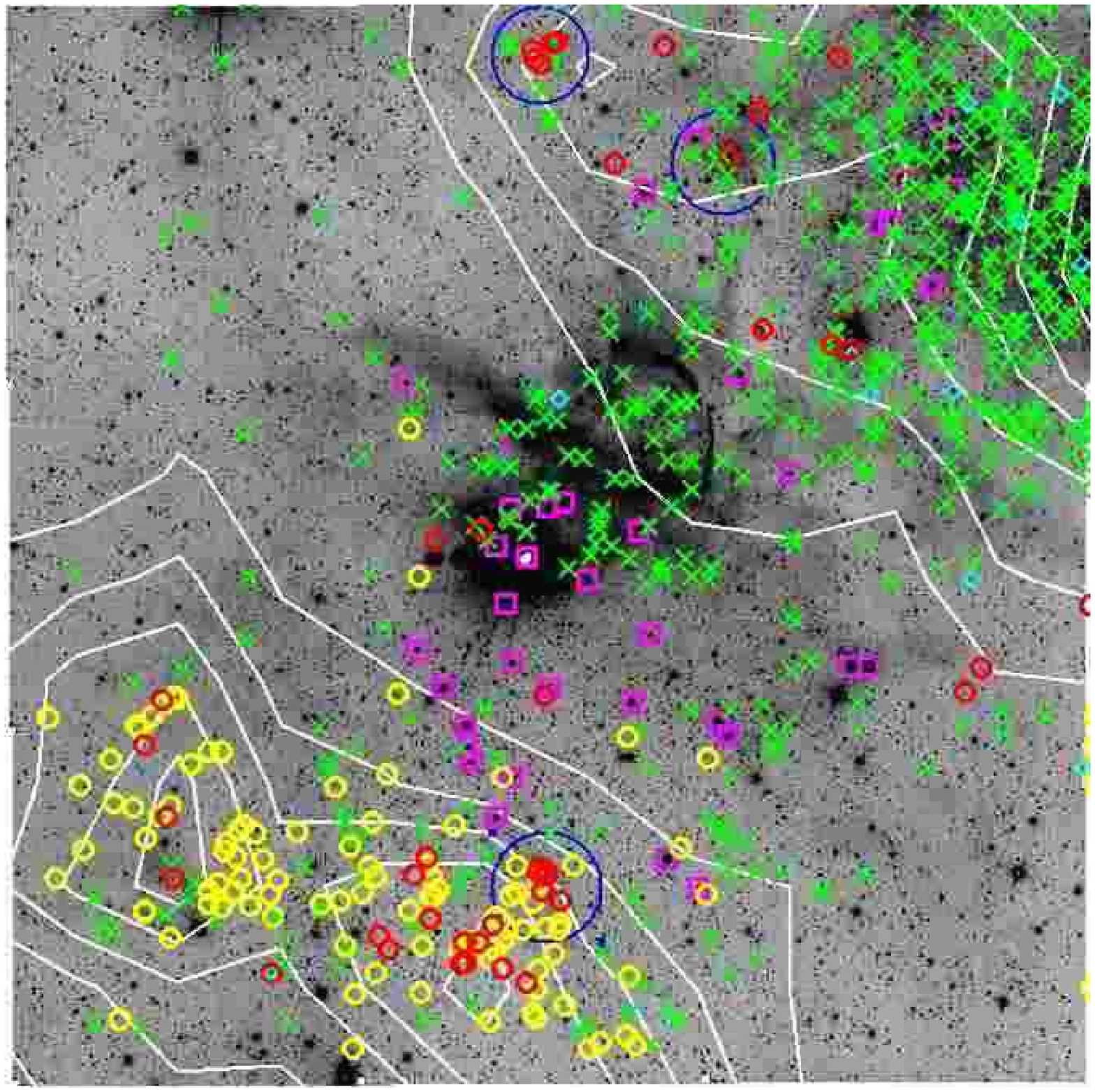}
\label{chandra}
\caption{The $K_s$ band mosaic image of the field, centered on $\eta$ Carinae (same as Fig.~1). The Class II and Class I candidates
identified from Fig. 2 are marked as crosses and diamonds, respectively. The sources cross-identified with known OB stars 
are marked as squares. Sources with $H-K_s>2$ are shown as dark circles, and sources detected only in the $K_s$ band with $K_s>17$ mag are shown as light circles. The white contours represent the $~^{12}$CO(1-0) emission \citep{brooks98}. The locations of the MSX point sources are marked with large circles.
} 
\end{center}
\end{figure}

\clearpage

\begin{figure}
\begin{center}
\plotone{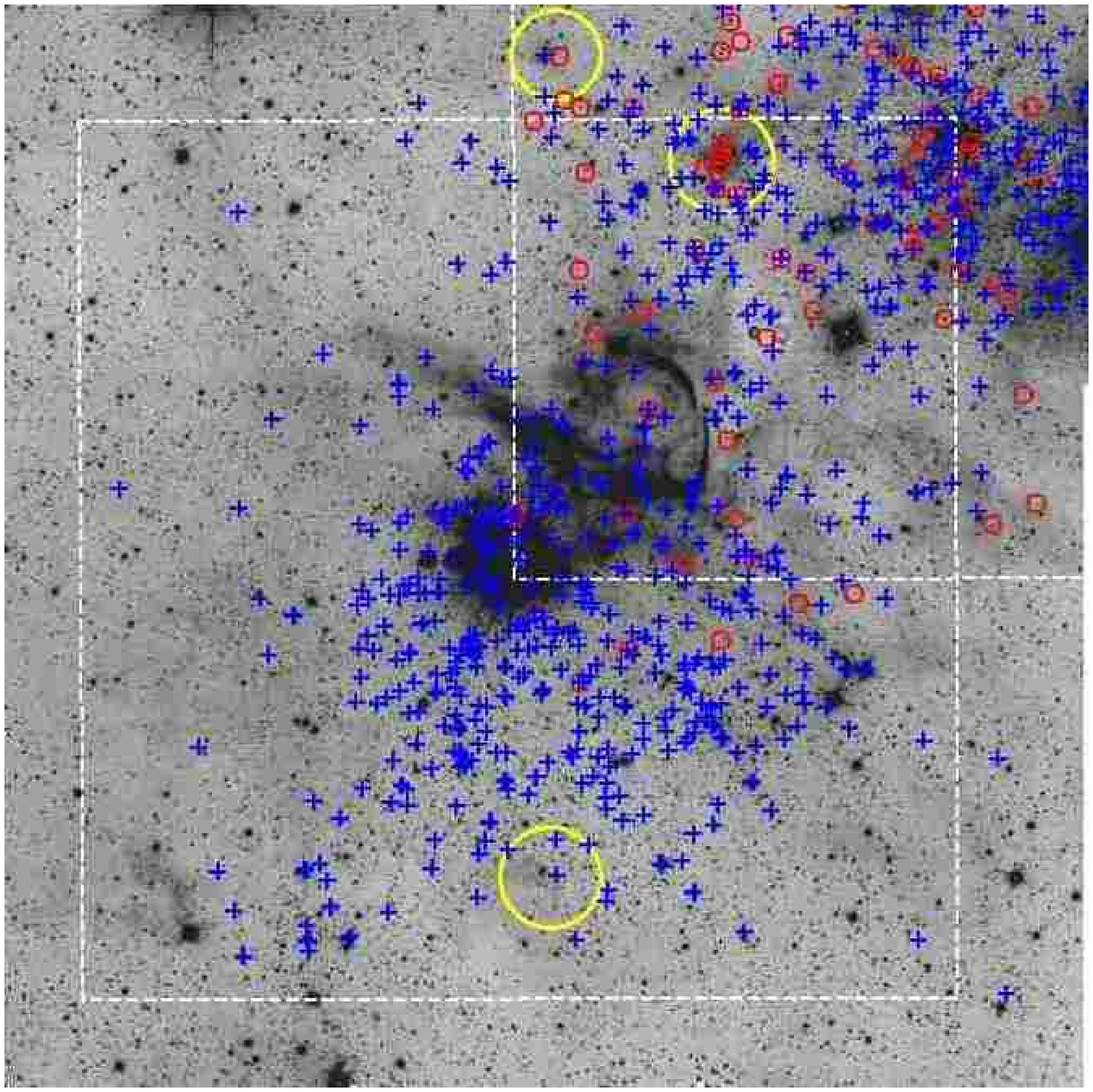}
\label{chandra}
\caption{The $K_s$ band mosaic image of the field, centered on $\eta$ Carinae (same as 
Fig.~1). The squares in the center and in the top show the fields covered by the $Chandra$ ObsIDs=1249,50 and ObsID=4495, respectively. The X-ray sources with and without an NIR counterpart are marked with pluses and circles, respectively. The locations of the MSX point sources are denoted with large circles.
} 
\end{center}
\end{figure}

\clearpage

\begin{figure}
\begin{center}
\plotone{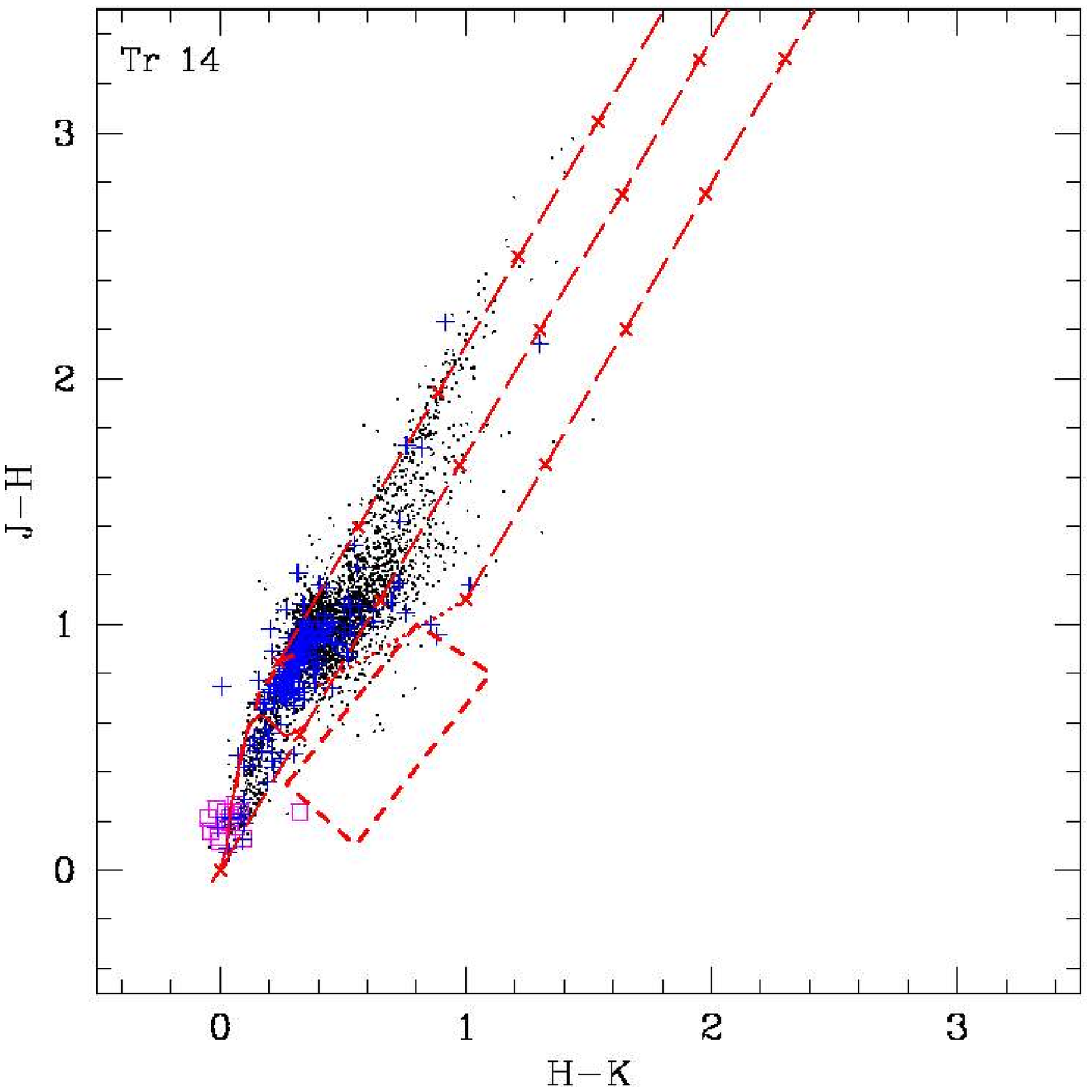}
\label{chandra}
\caption{The color-color diagram of the Tr\,14 region. The dwarf and giant loci, the reddening band, the CTTS locus, and the region for Herbig Ae/Be stars, are plotted in a similar way as in Fig.~2. The known O or early-B stars are shown as open squares, whereas the NIR sources with X-ray counterparts are shown as plus symbols.
} 
\end{center}
\end{figure}

\clearpage

\begin{figure}
\begin{center}
\plotone{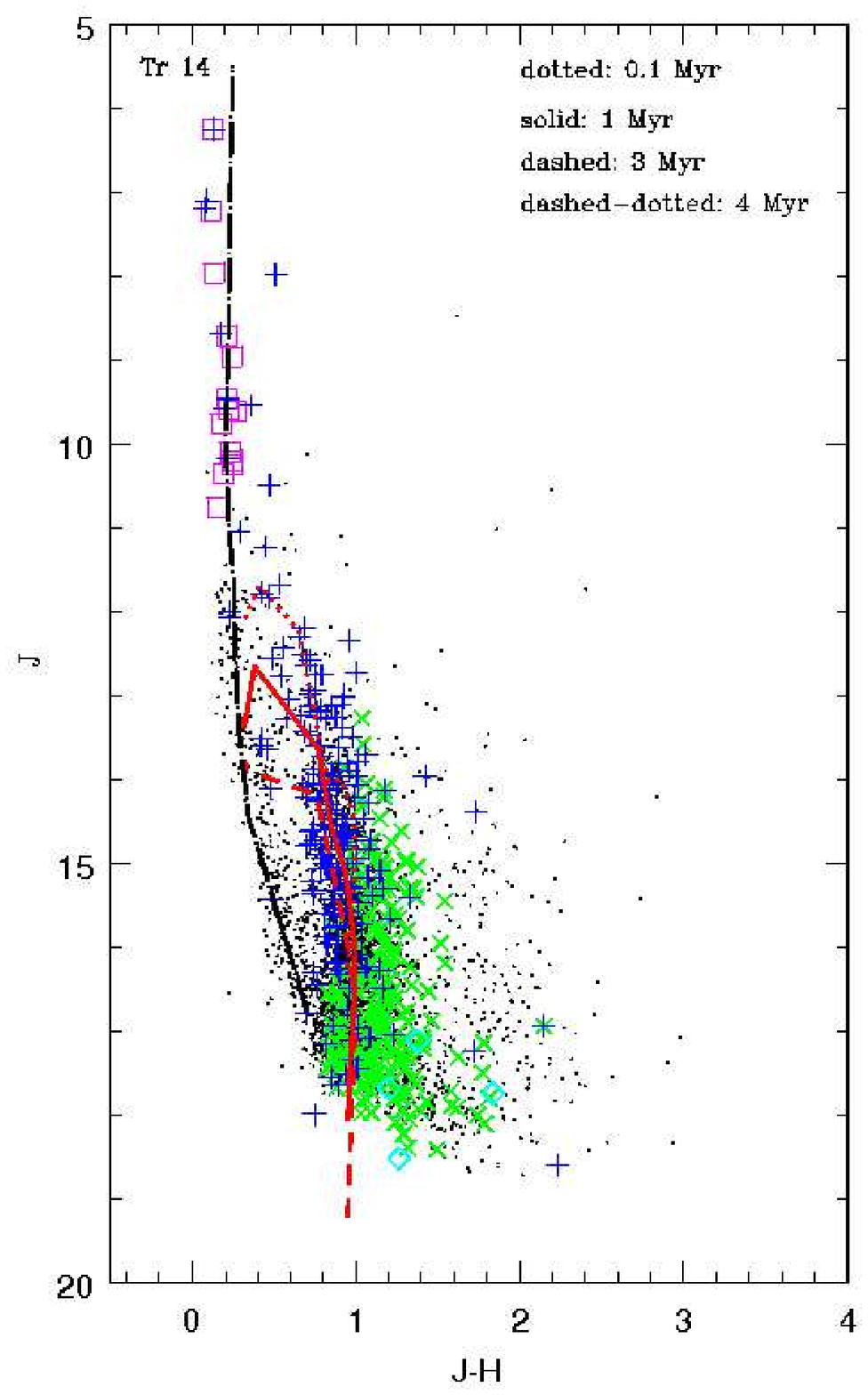}
\label{chandra}
\caption{The color-magnitude diagram of the Tr\,14 region. The known O or early-B stars are shown as open squares, the CTTS candidates selected from Fig.~7 are shown as crosses, whereas the NIR sources 
with X-ray counterparts are shown as pluses. The dashed dotted line represents 4 Myr post--main-sequence evolutionary track
\citep{bertelli}, whereas the dotted line, the solid line, and the dashed line represent the 0.3 Myr, 1 Myr and 3 Myr PMS evolutionary tracks, respectively \citep{palla99}.
} 
\end{center}
\end{figure}

\clearpage

\begin{figure}
\begin{center}
\plotone{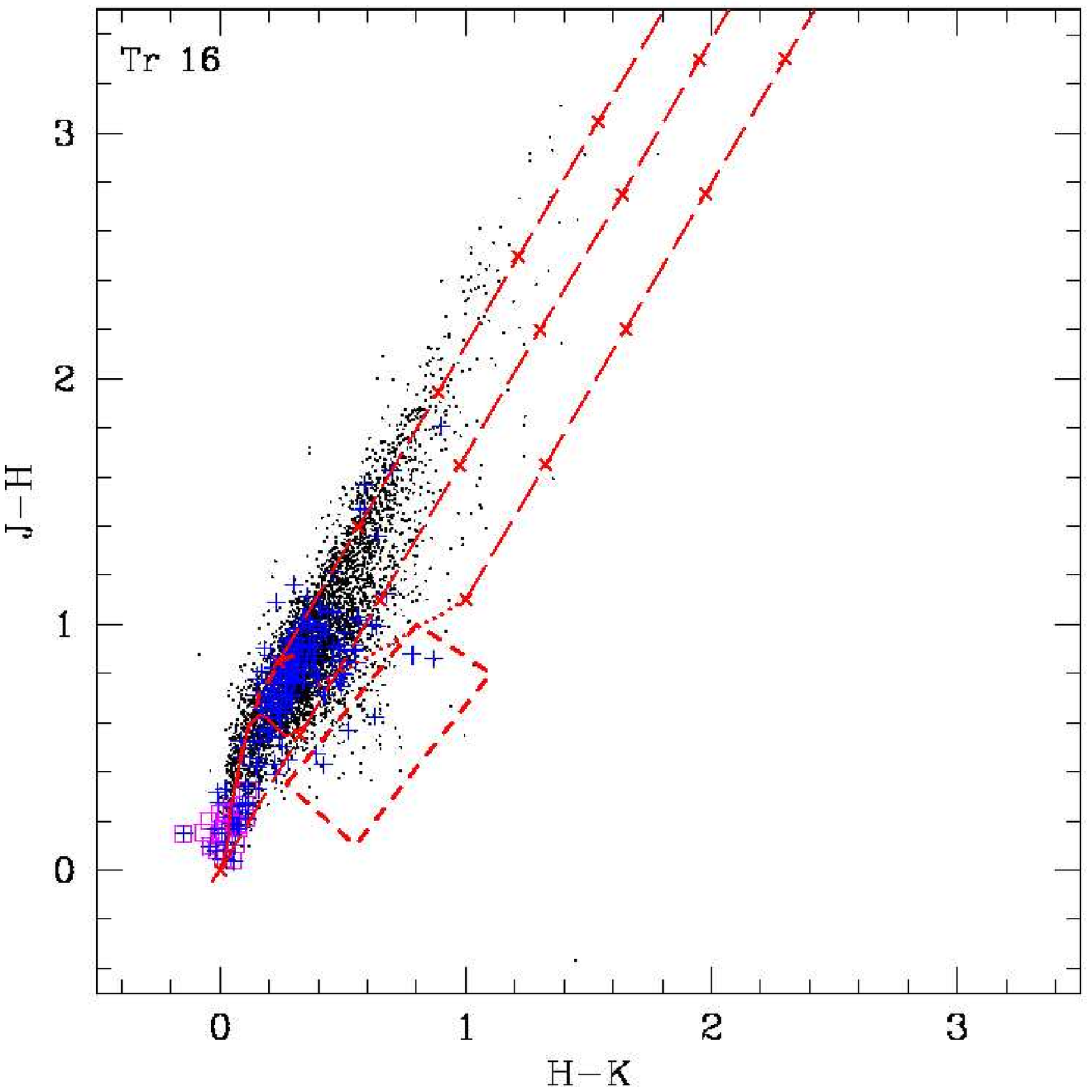}
\label{chandra}
\caption{The same as Fig.~7 except for Tr\,16
} 
\end{center}
\end{figure}

\clearpage

\begin{figure}
\begin{center}
\plotone{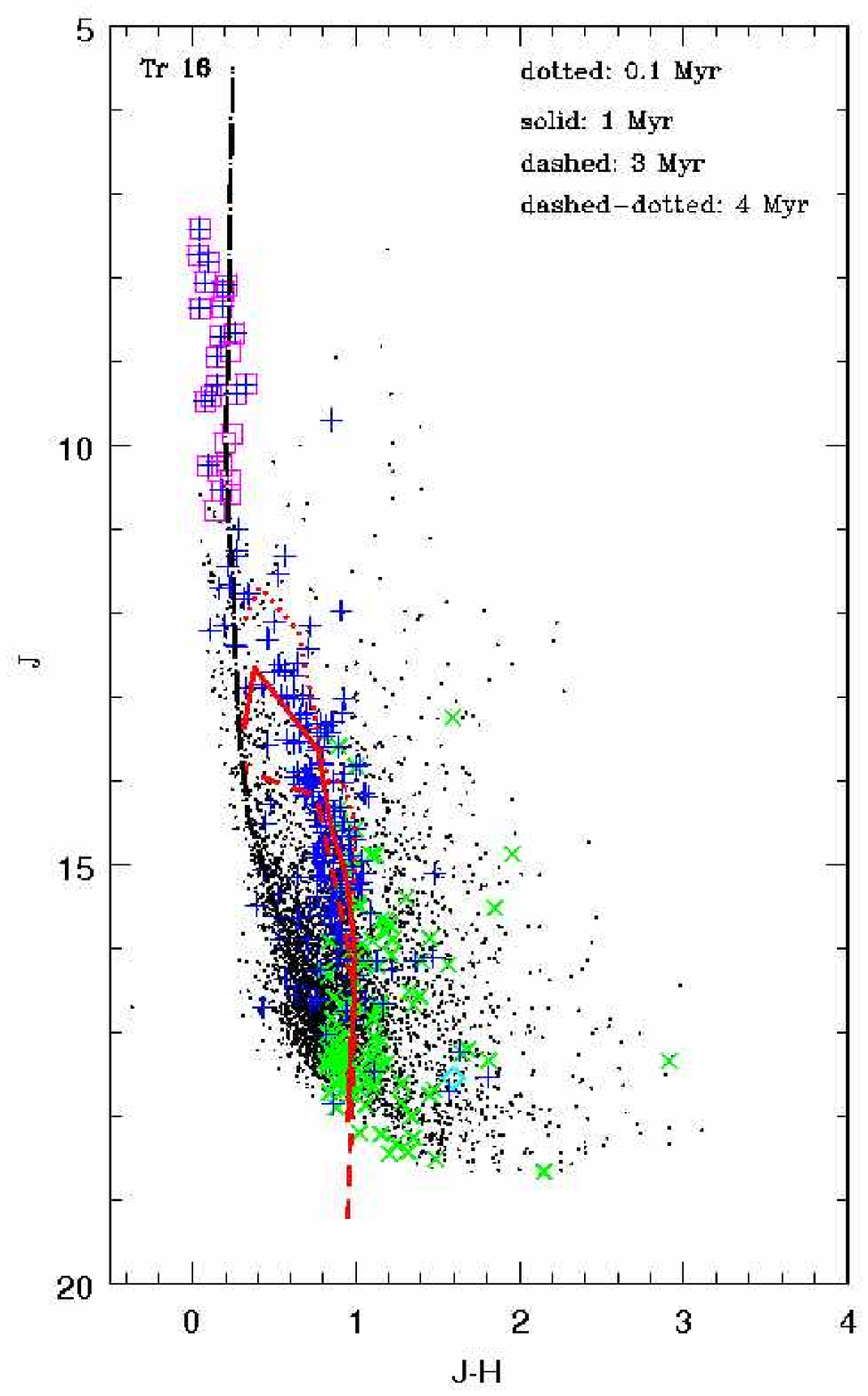}
\label{chandra}
\caption{The same as Fig.~8 except for Tr\,16.
} 
\end{center}
\end{figure}

\clearpage

\begin{figure}
\begin{center}
\plotone{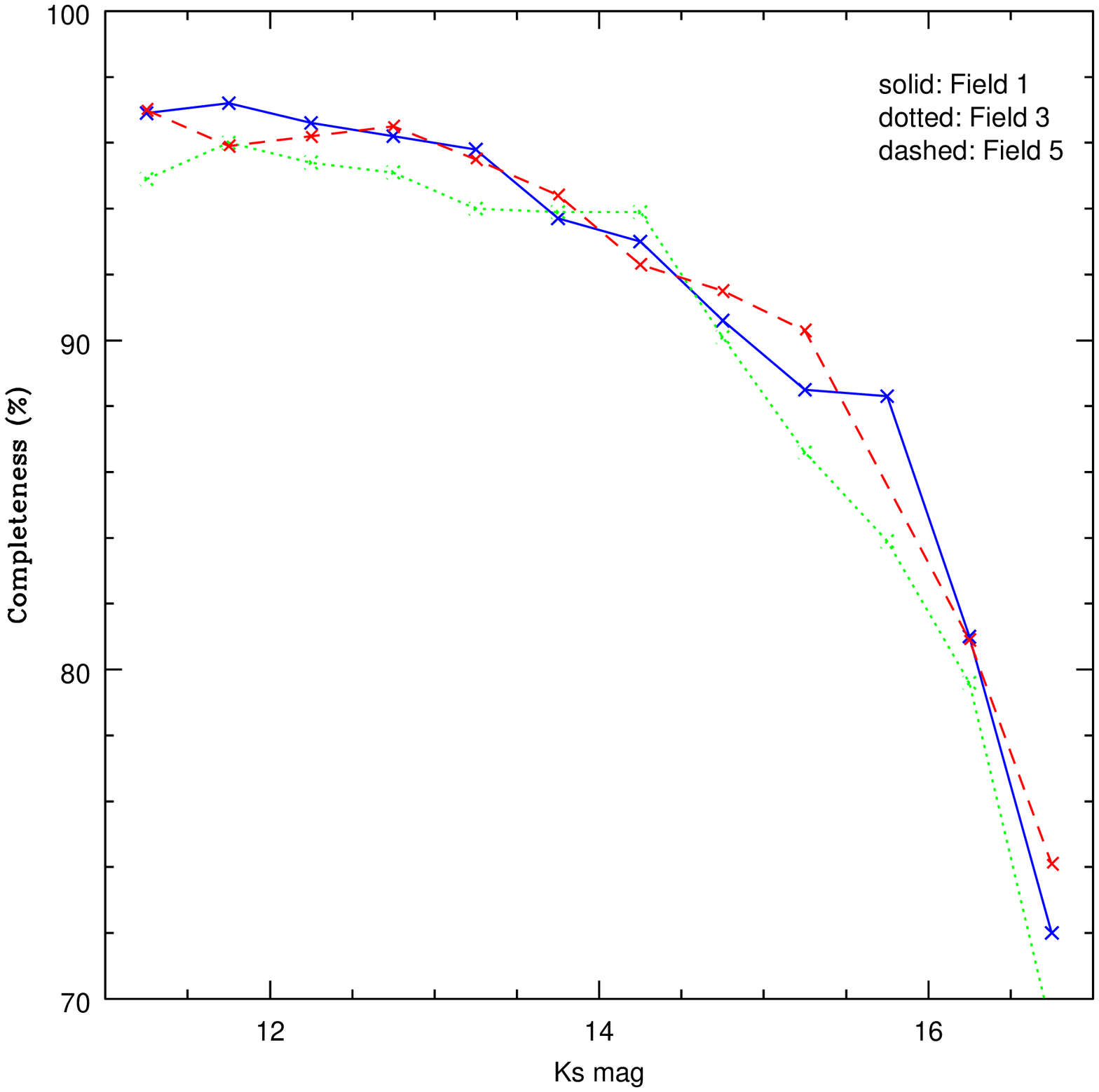}
\caption{The completeness of the $K_s$-band data for fields, No. 1 (solid line), No. 3 (dotted line), and No. 5 (dashed line). Field No. 3 includes Tr\,14 and field No. 5 includes Tr\,16.}
\end{center}
\end{figure}



\clearpage

\begin{figure}
\begin{center}

\vspace{-3.0cm}

\plotone{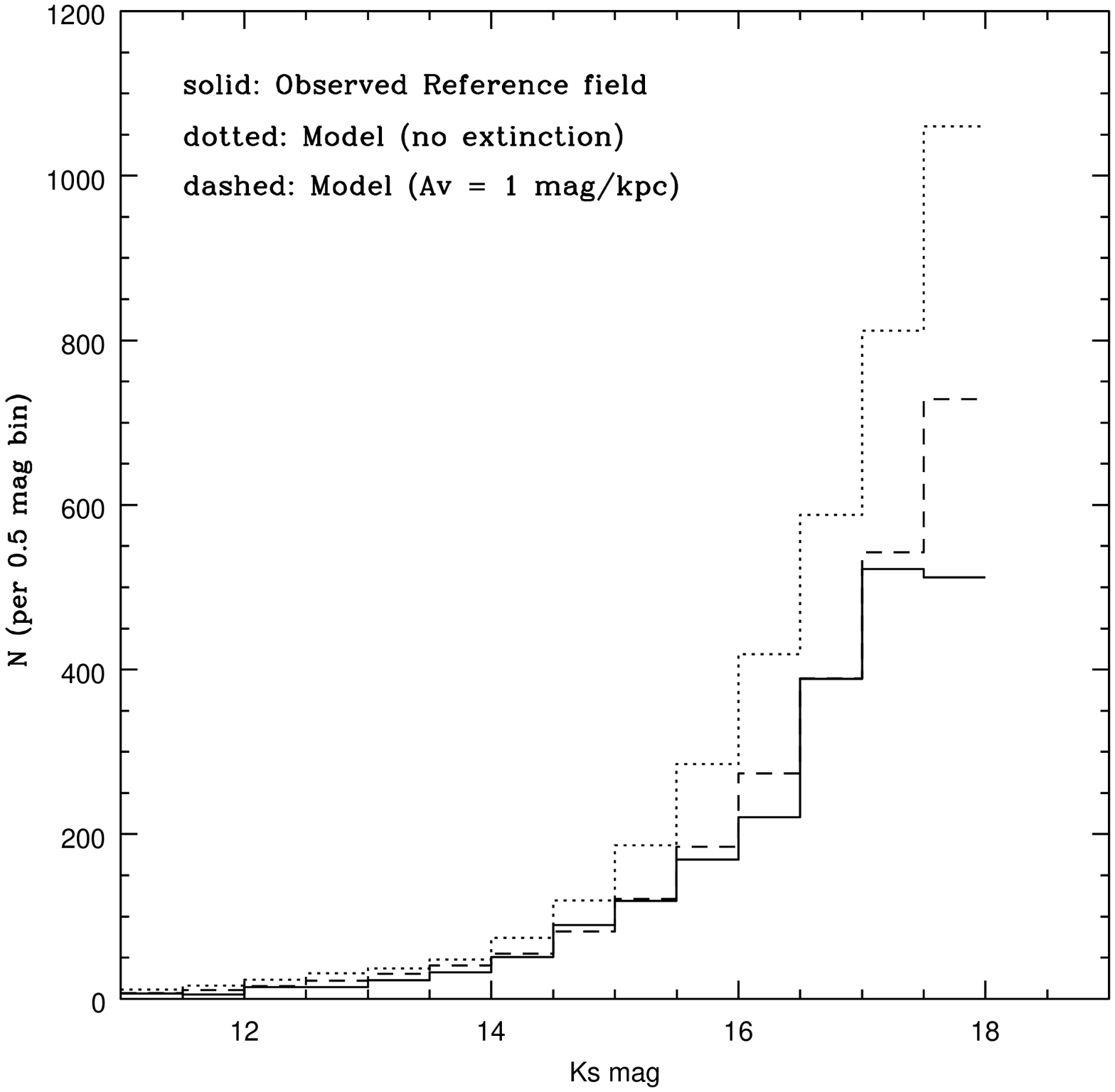}
\label{chandra}
\caption{The completeness corrected KLF of the reference field (solid line) is compared with the model predicted KLF when no extinction is applied (dotted line) and with the model predicted KLF where an extinction of $A_V$ = 1 mag~kpc$^{-1}$ is applied to all the model stars (dashed line).
} 
\end{center}
\end{figure}

\clearpage

\begin{figure}
\begin{center}
\includegraphics[scale=0.45]{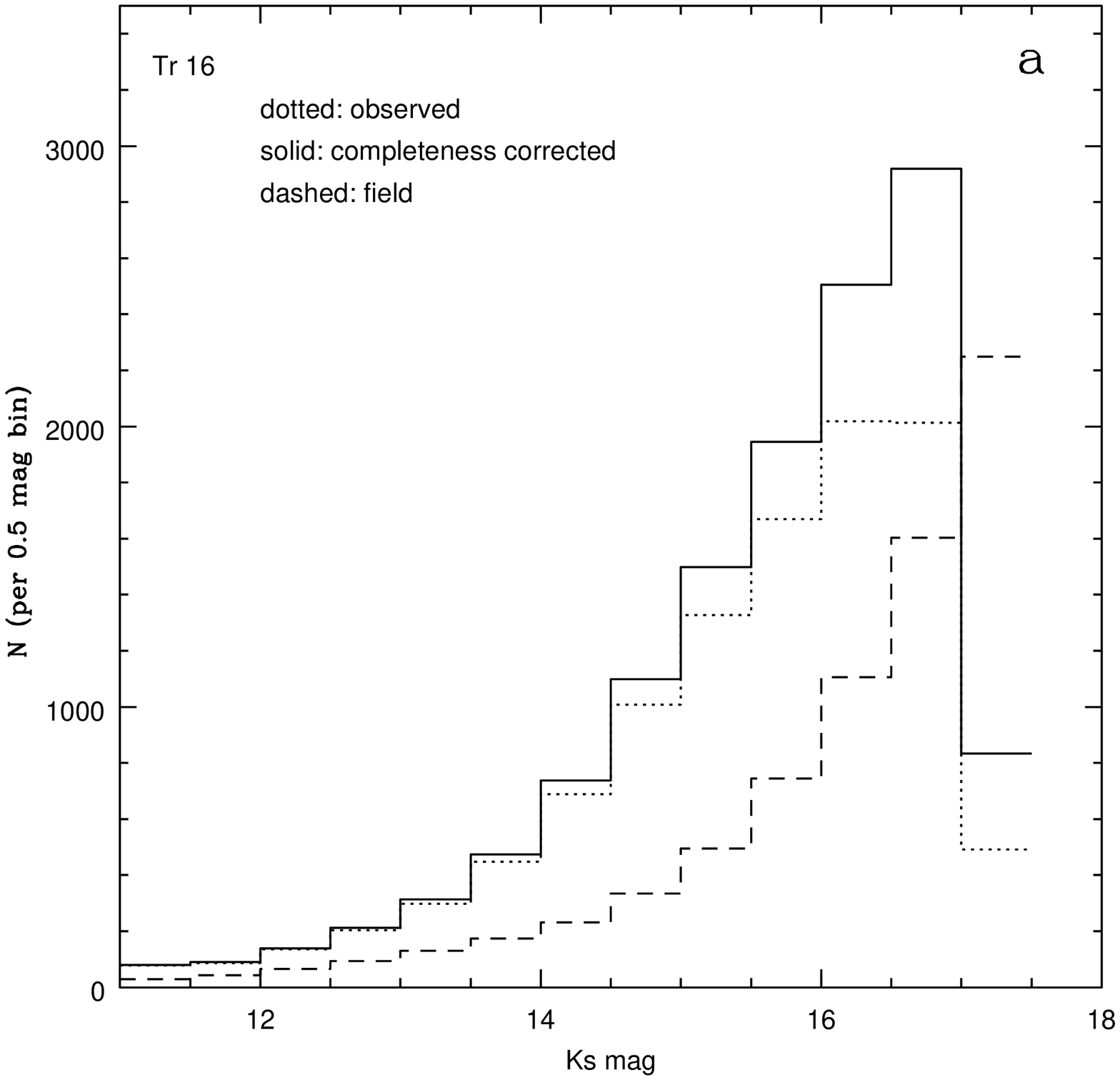}

\includegraphics[scale=0.45]{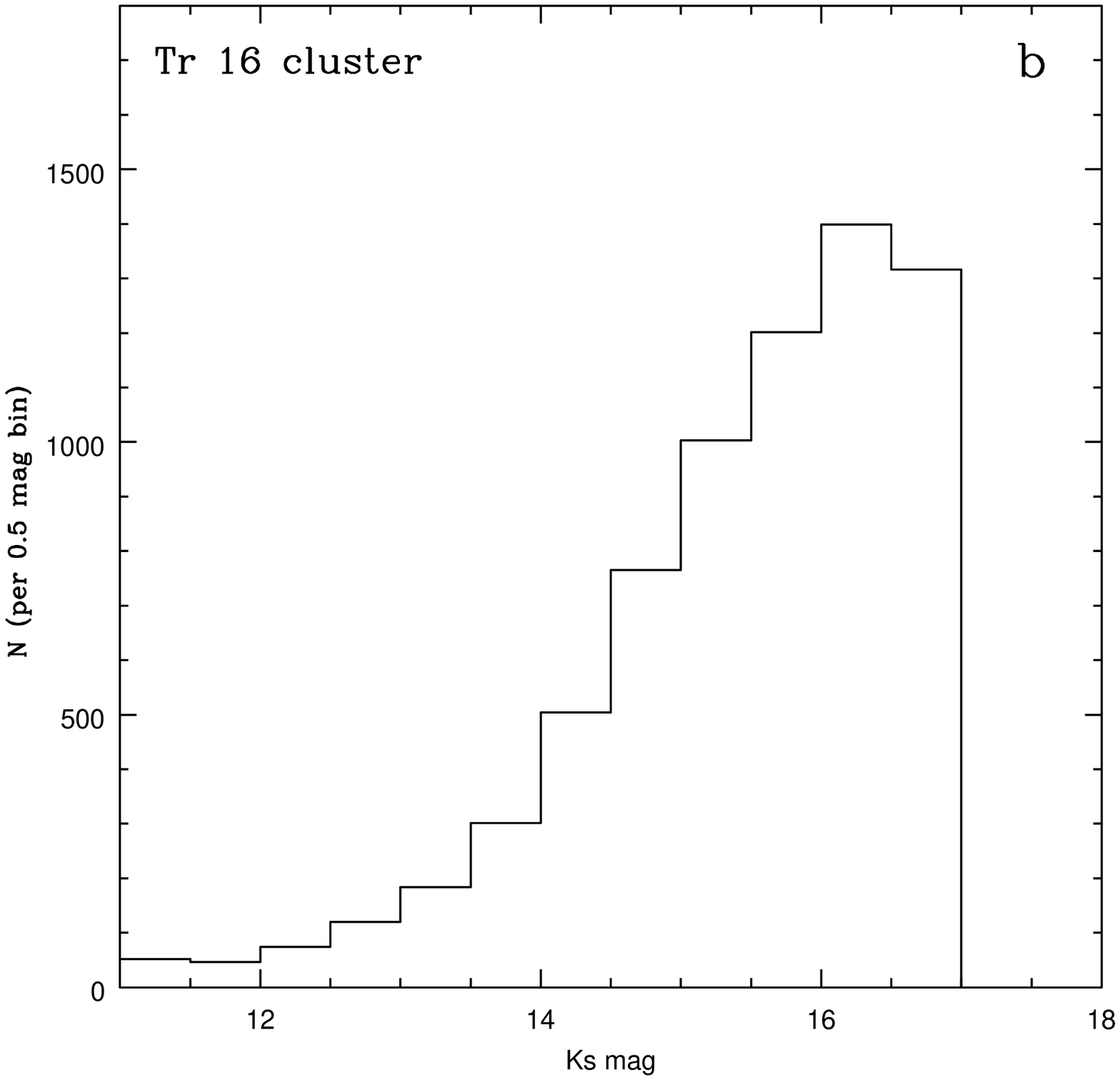}

\label{chandra}
\caption{(a) The observed KLF of Tr\,16 (dotted line) with the completeness corrected KLF of Tr\,16 (solid line) and the field star contribution (dashed line). The field star contribution is computed using \citet{robin03} model by applying an interstellar extinction of $A_V$ = 1 mag~kpc$^{-1}$ to all the stars and an additional cloud extinction of $A_V$ =1.5 mag to the background stars. (b) The cluster KLF of Tr\,16, where the field star contribution has been subtracted from the completeness corrected KLF.} 
\end{center}
\end{figure}

\clearpage

\begin{figure}
\begin{center}
\includegraphics[scale=0.45]{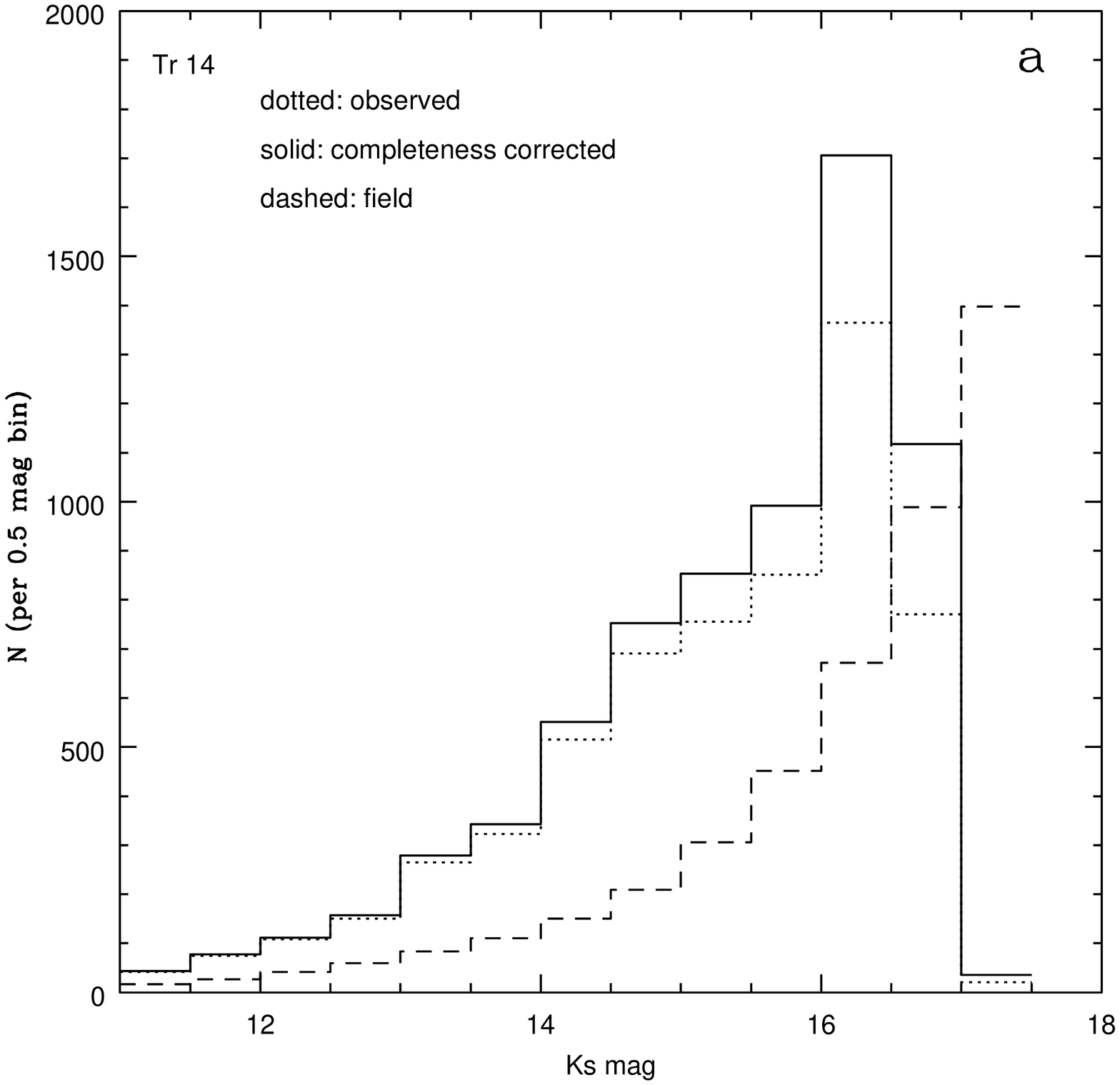}

\includegraphics[scale=0.45]{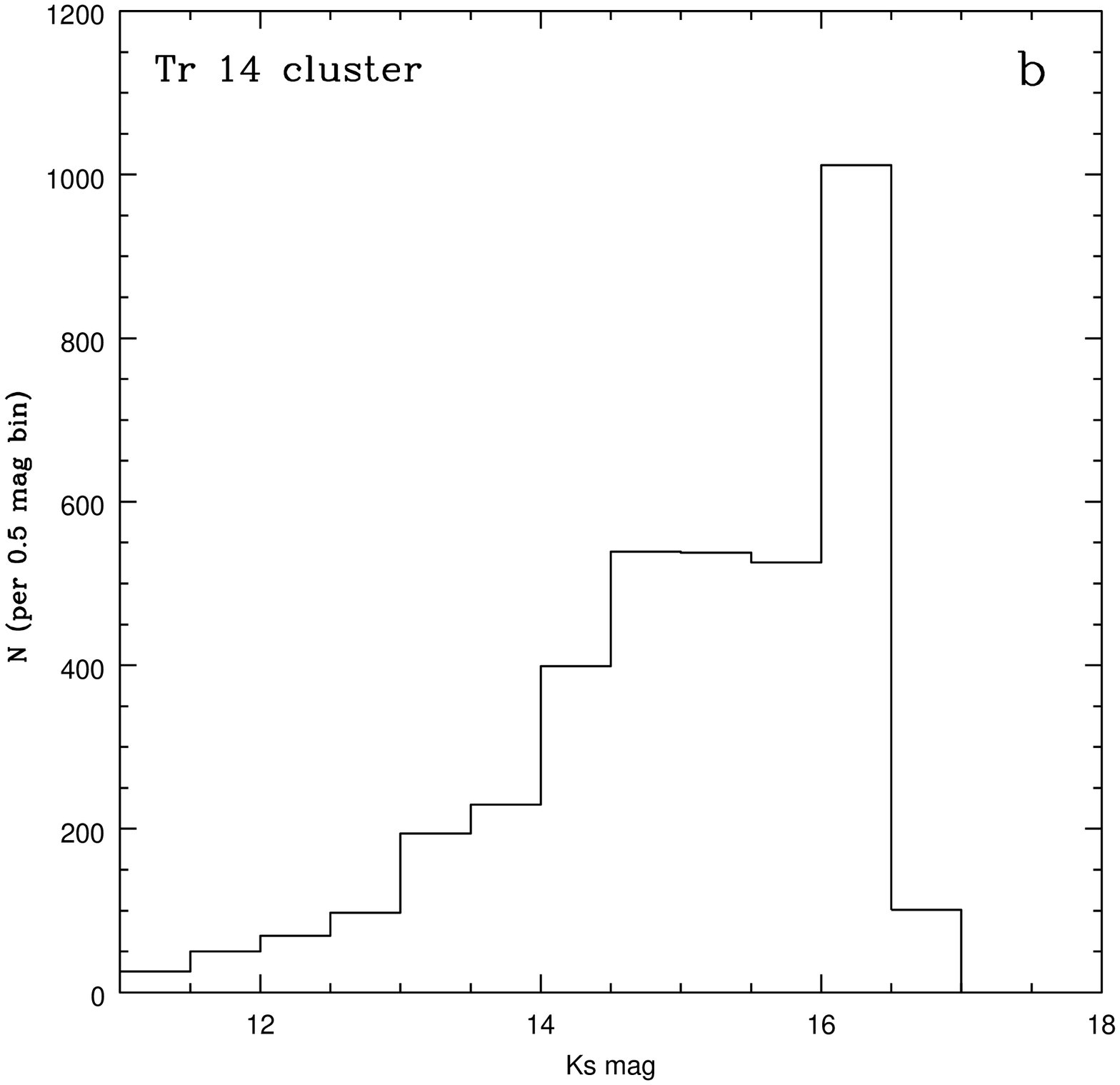}
\label{chandra}
\caption{The same as Fig.~13 except for Tr\,14, for which the additional cloud extinction of $A_V$ =2.5 mag is applied to the background stars.} 
\end{center}
\end{figure}




\clearpage

\begin{figure}
\begin{center}
\includegraphics[scale=0.45]{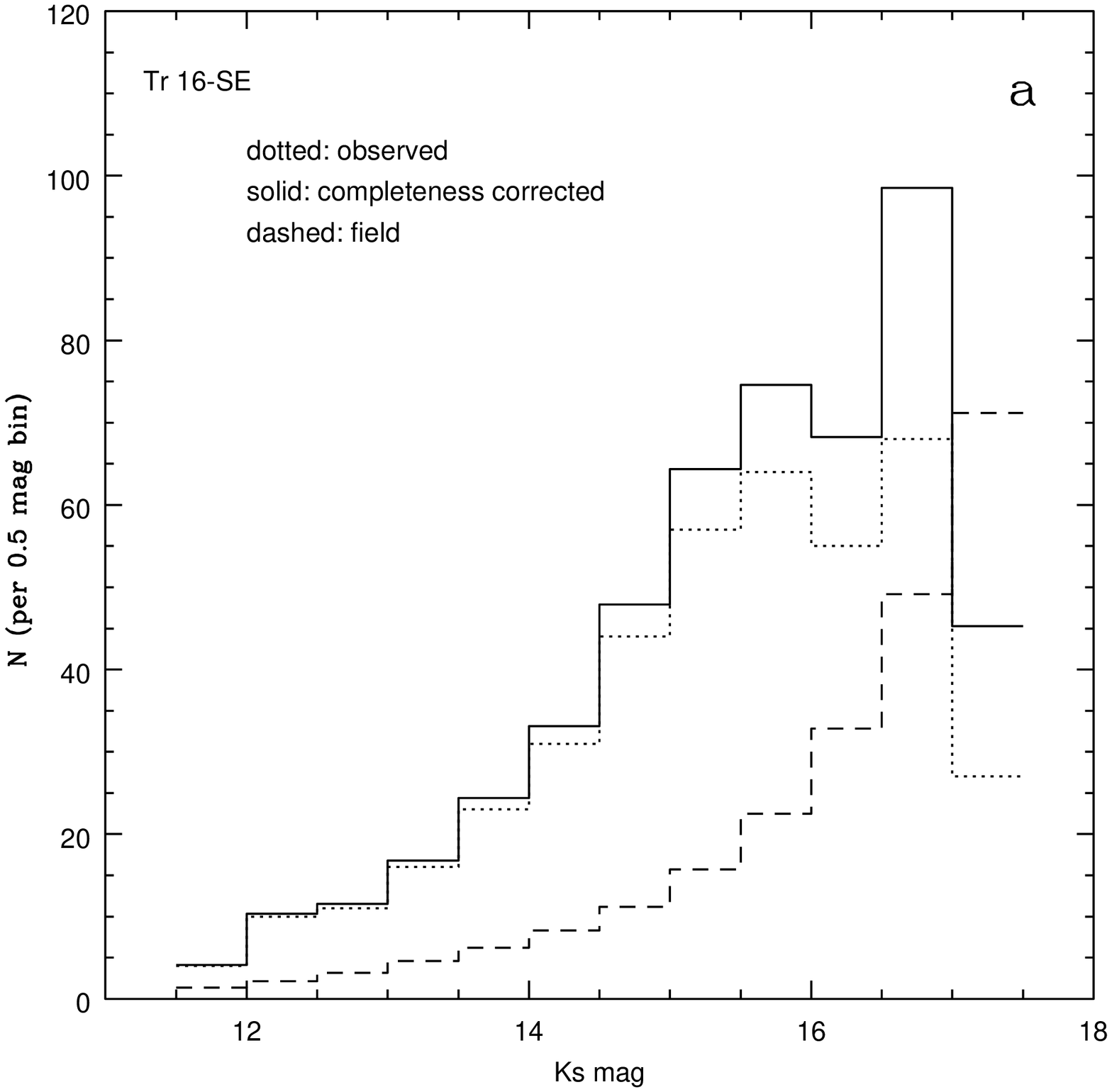}

\includegraphics[scale=0.45]{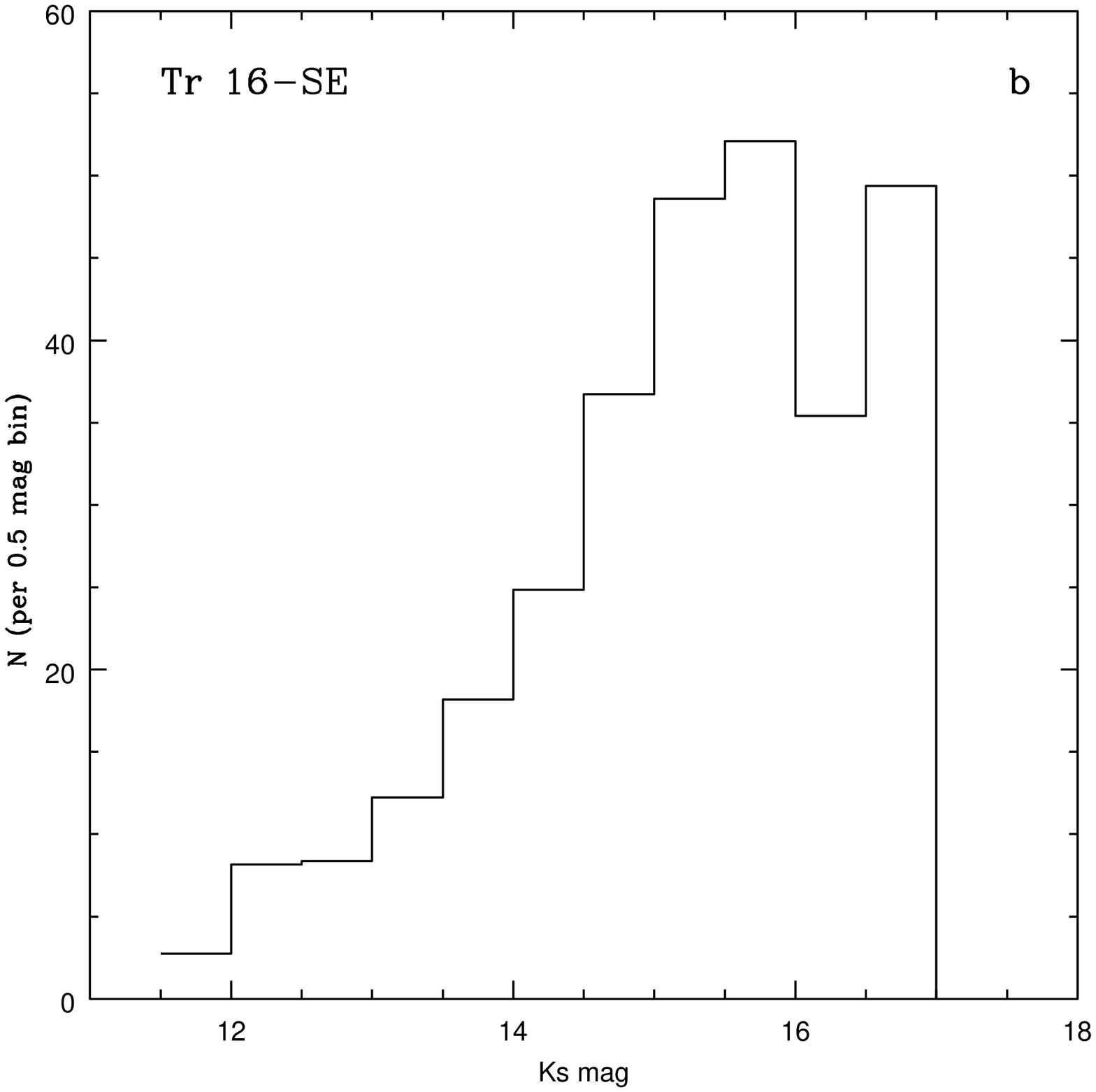}
\label{chandra}
\caption{The same as Fig.~13 except for the Tr\,16-SE group, for which the additional cloud extinction of $A_V$ =5.5 mag is applied to the background stars.} 
\end{center}
\end{figure}

\clearpage

\begin{figure}
\begin{center}

\vspace{-3.0cm}

\includegraphics[scale=0.5]{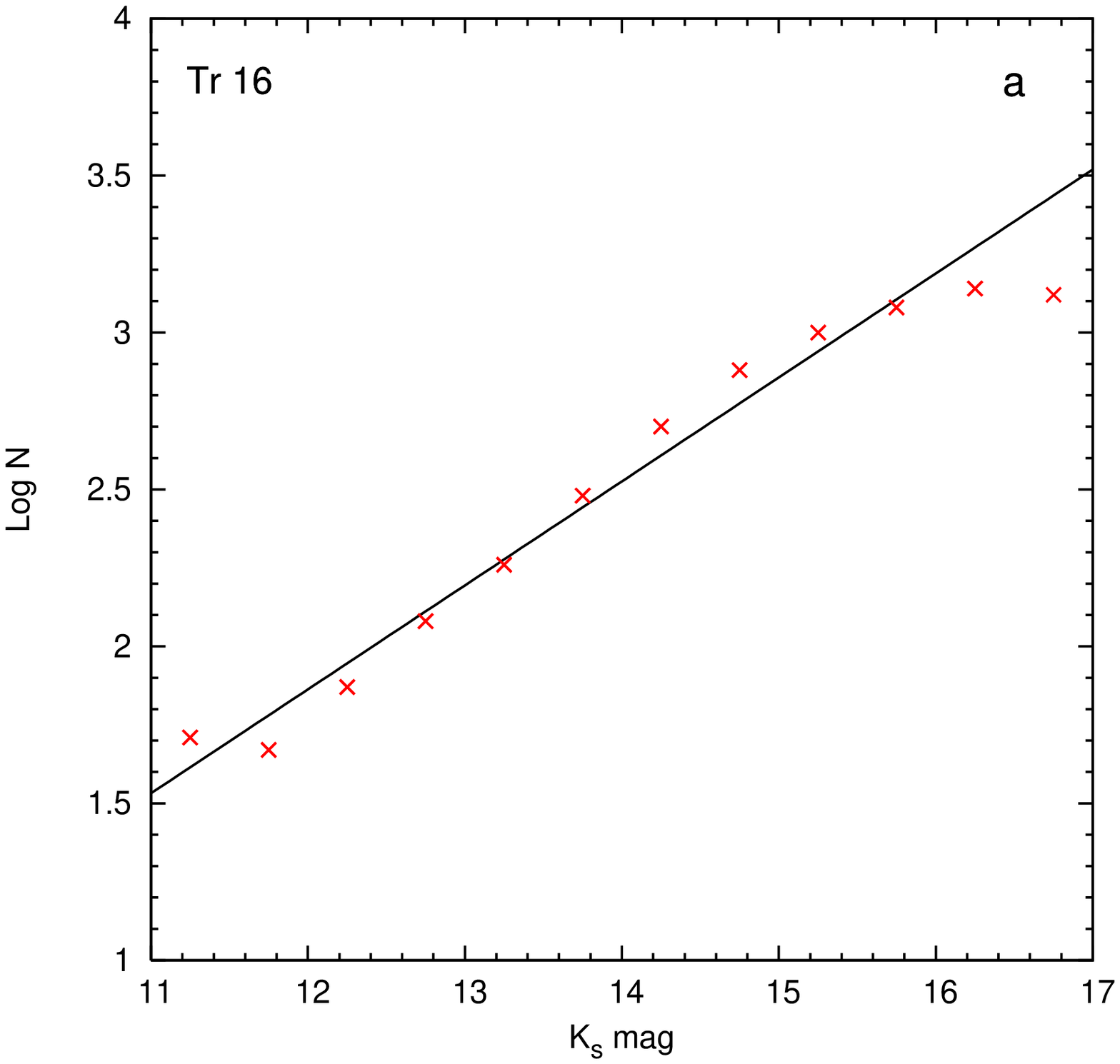}

\vspace{-3.0cm}

\includegraphics[scale=0.45]{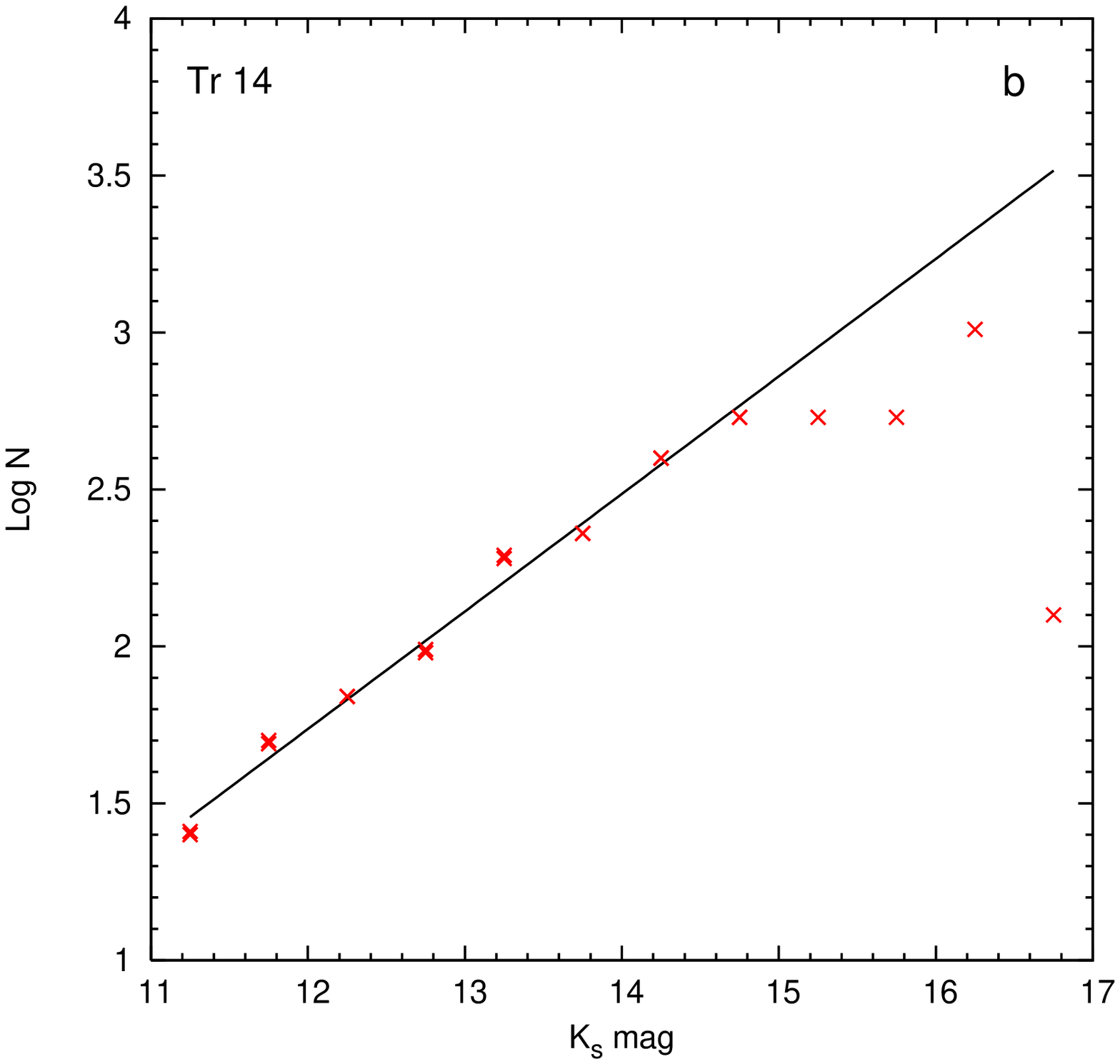}
\includegraphics[scale=0.45]{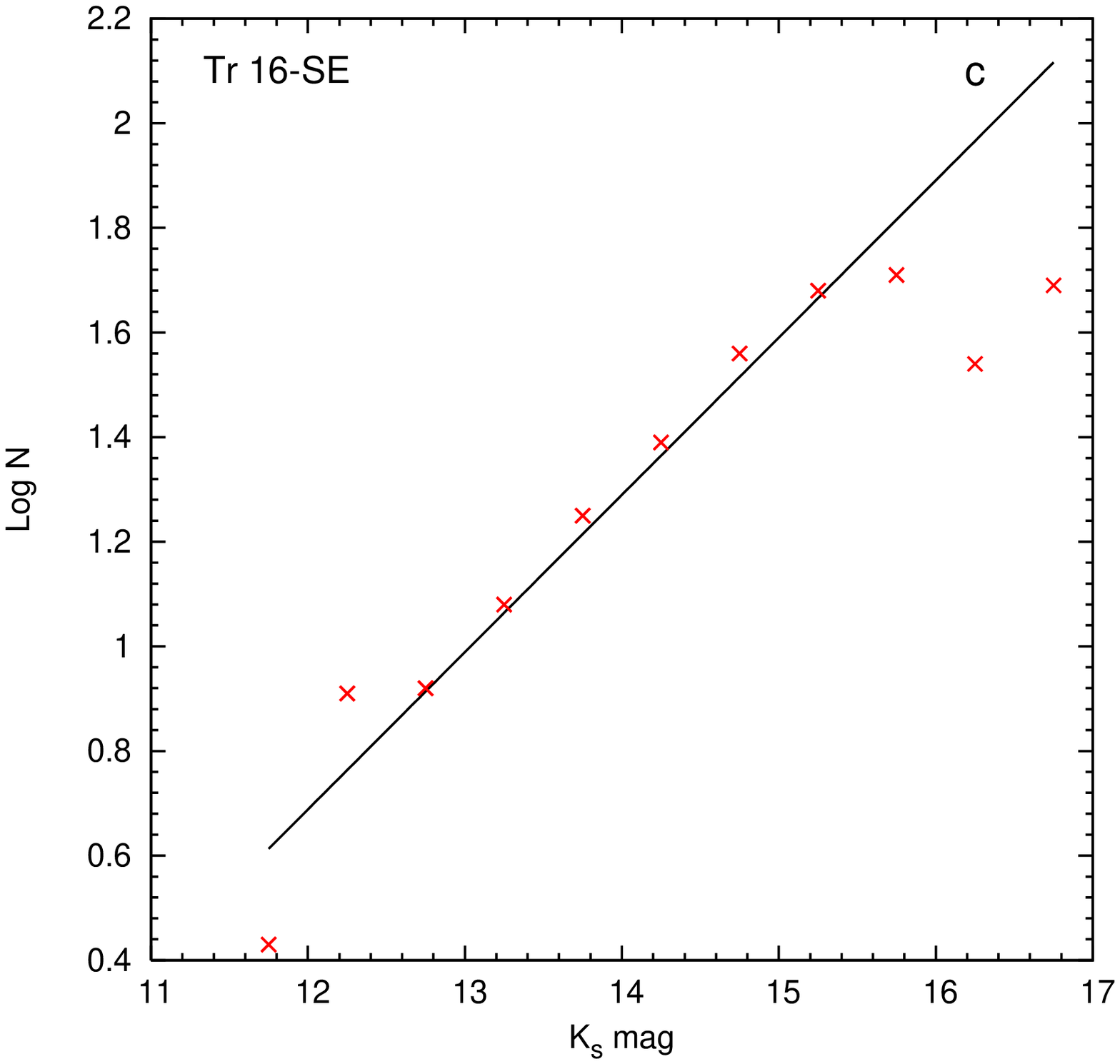}
\label{chandra}
\caption{The cluster KLFs of (a) Tr\,16, (b) Tr\,14, and (c) Tr\,16-SE group. The total number of $K_s$ sources in the cluster KLFs are, 6961, 3773, and 292 for Tr\,16, Tr\,14, and Tr\,16-SE, respectively. A power-law with a slope $\alpha$ is fitted to each KLF until the KLF starts to turnover, where $\alpha$ = 0.33, 0.37, and 0.30, for Tr\,16, Tr\,14, and Tr\,16-SE group, respectively.} 
\end{center}
\end{figure}

\clearpage

\begin{figure}
\begin{center}

\vspace{-3.0cm}

\includegraphics[scale=0.5]{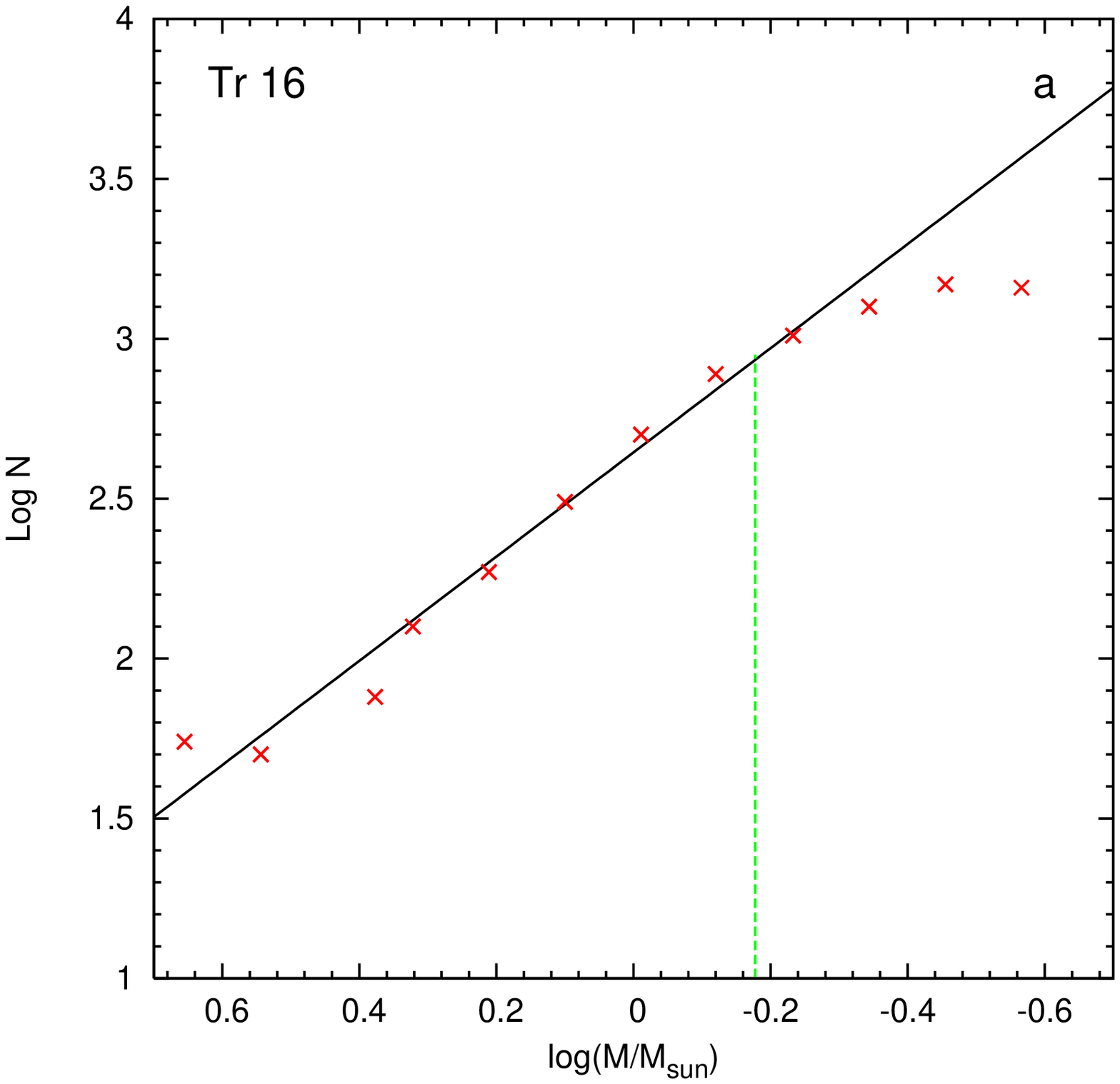}

\vspace{-3.0cm}

\includegraphics[scale=0.45]{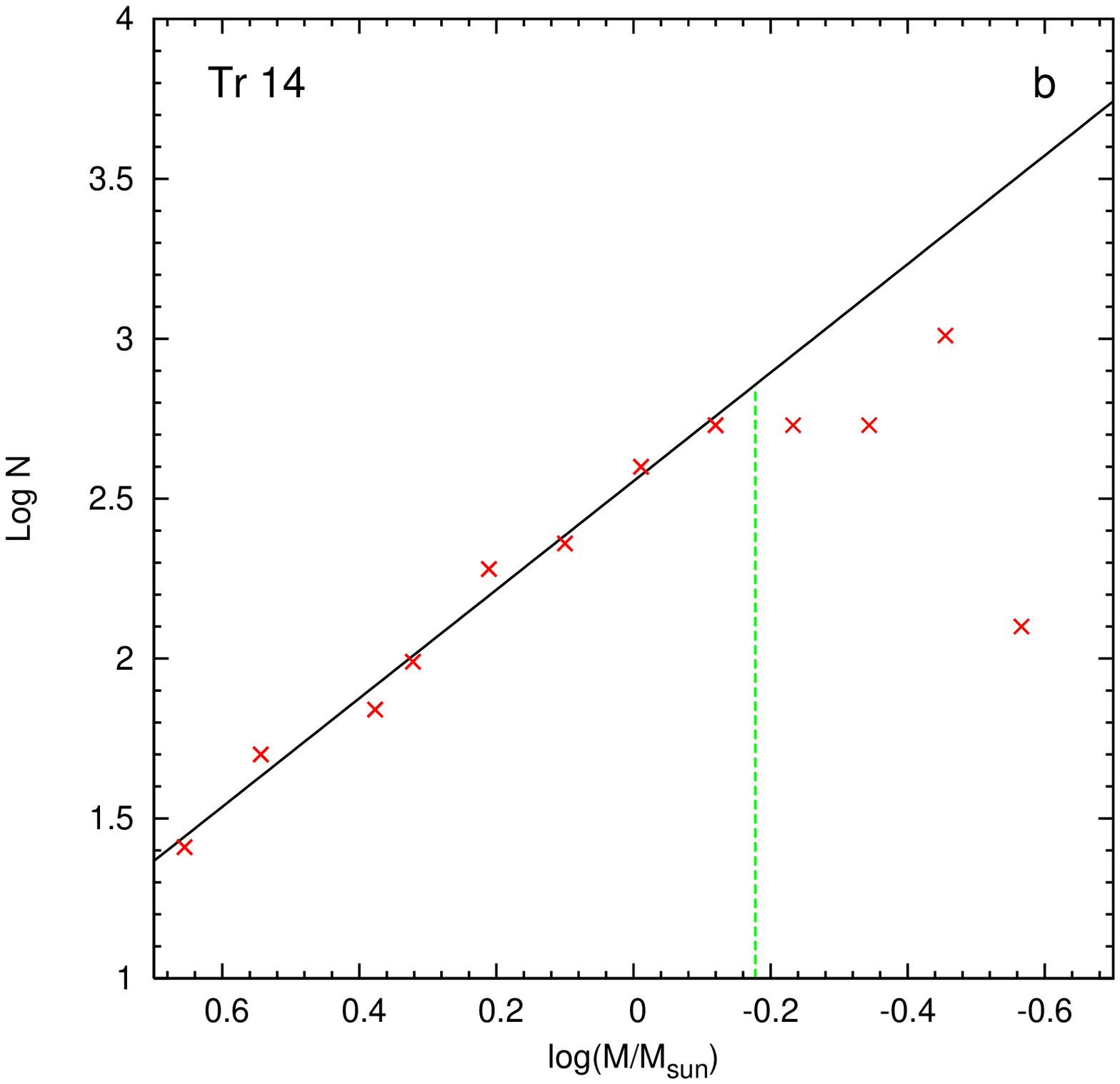}
\includegraphics[scale=0.45]{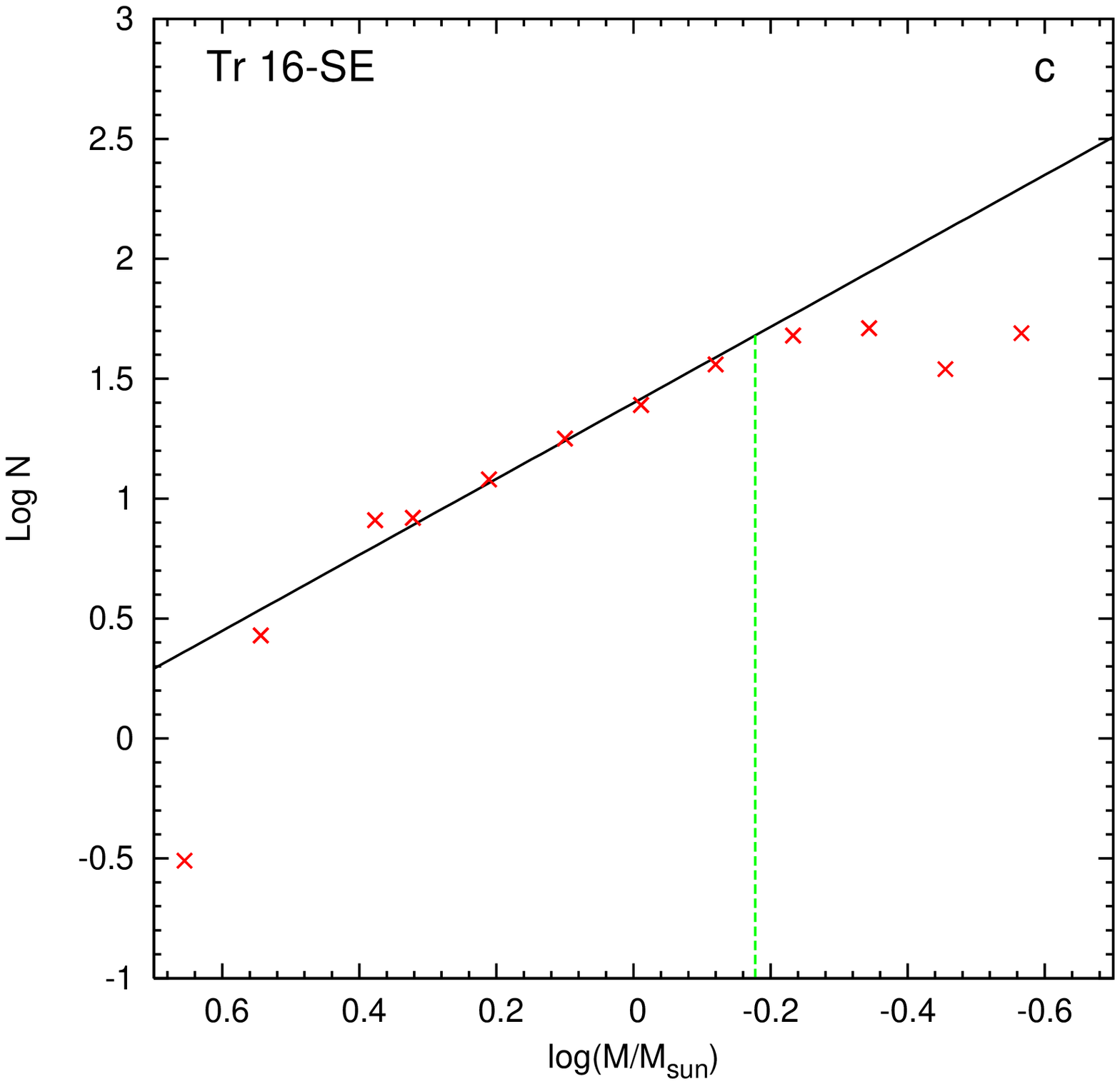}
\label{chandra}
\caption{Mass functions of (a) Tr\,16, (b) Tr\,14, and (c) Tr\,16-SE group, derived from the cluster KLFs. The slopes of the mass functions of Tr\,16, Tr\,14 and Tr\,16-SE group
are found to be $-$1.62, $-$1.69, and $-$1.58 respectively, until the 90\% completeness limit (dashed line), i.e., $\sim$0.7 \Msun.} 
\end{center}
\end{figure}

\clearpage

\begin{figure}
\begin{center}
\includegraphics[scale=0.40]{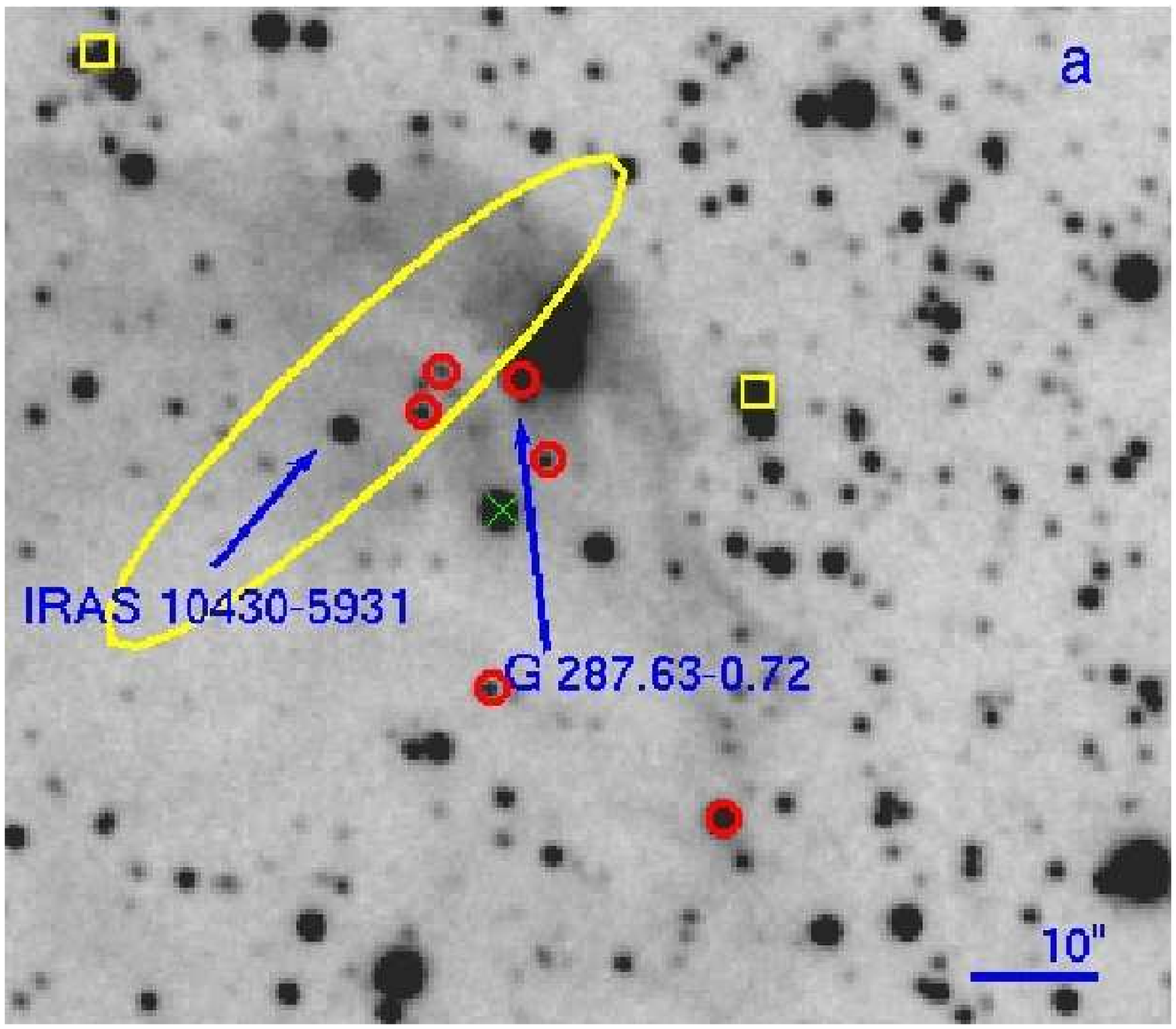}
\includegraphics[scale=0.42]{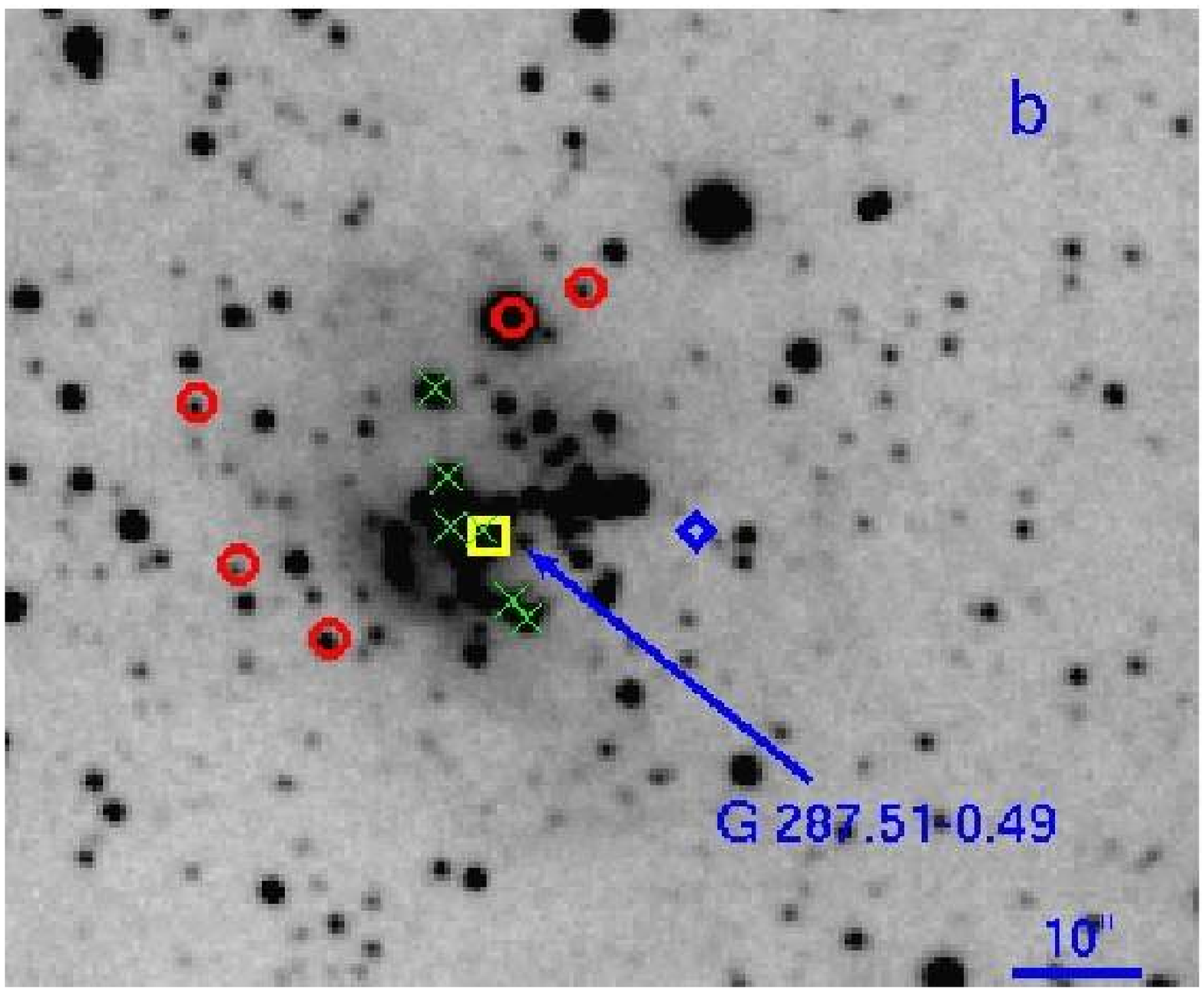}

\includegraphics[scale=0.45]{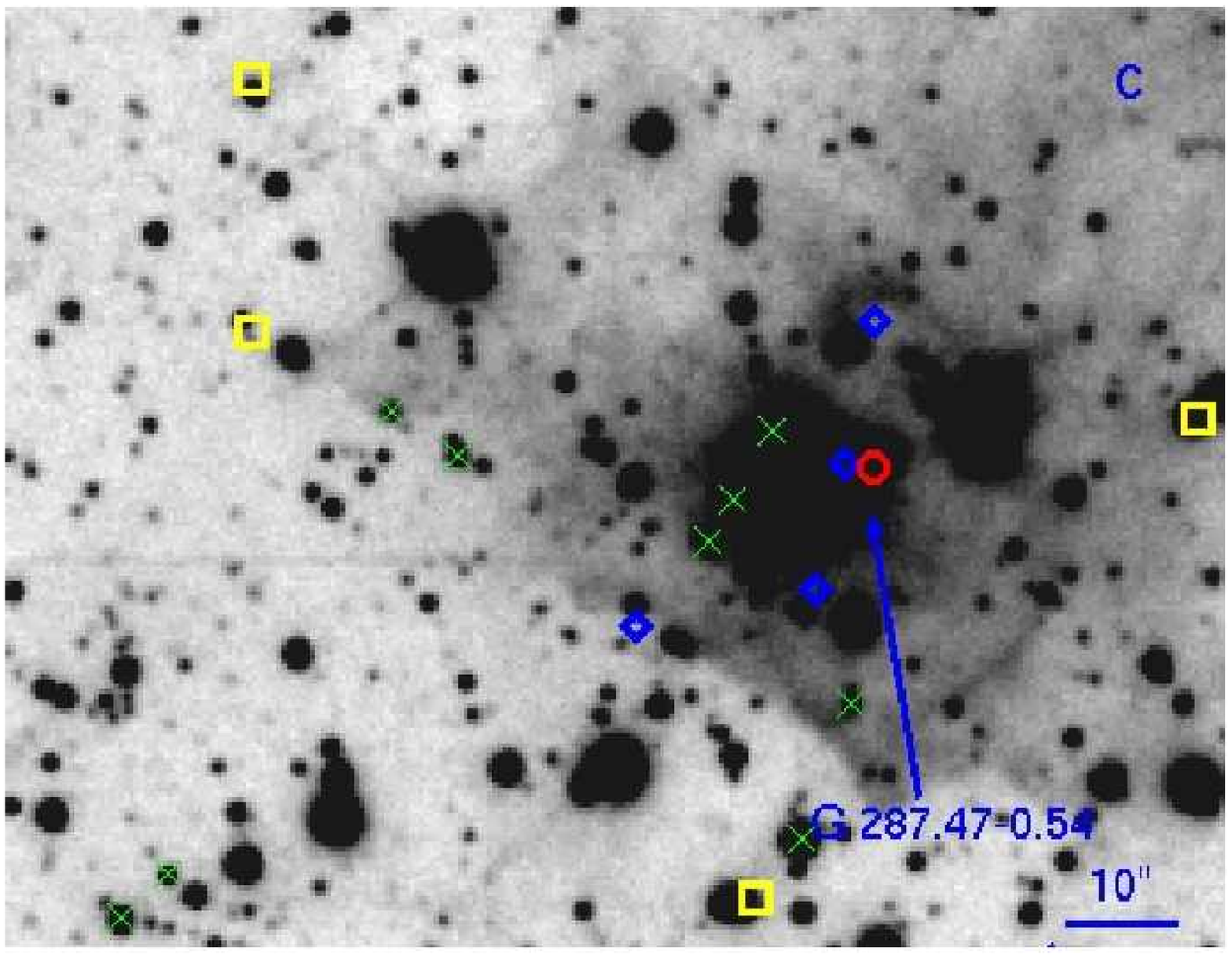}

\includegraphics[scale=0.45]{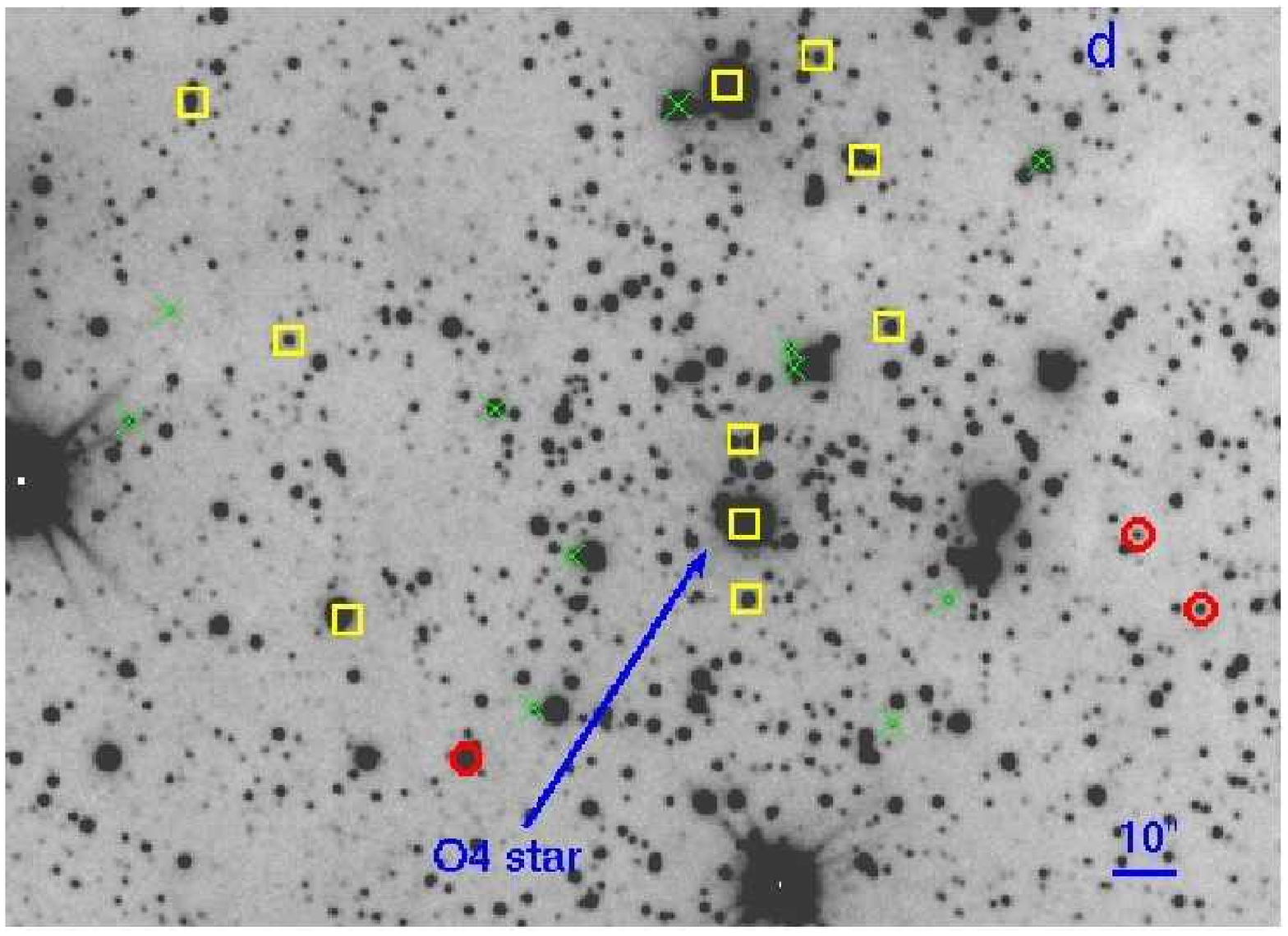}
\end{center}
\caption{The $K_s$ band images around (a) IRAS\,10430-5931, (b) 
G\,287.51-0.49, (c) G\,287.47-0.54, and (d) Tr\,16-SE group. The IRAS and MSX sources are marked by arrows and labeled. Positional uncertainties (ellipse) for the IRAS source are marked on the image. Also marked are the T Tauri candidates (crosses), red sources with $H-K_s>2$ (circles), X-ray sources with NIR counterparts (boxes), and X-ray sources with no counterparts (diamonds).}
\end{figure}


\end{document}